\documentclass[useAMS,usenatbib]{mn2e} 
\usepackage{aas_macros}
\usepackage{graphics}
\usepackage[pdftex]{graphicx}
\usepackage{epstopdf}
\usepackage{epsfig}  
\usepackage{natbib} 
\usepackage{float}
\usepackage{amsmath}
\usepackage{times}
\usepackage[varg]{txfonts}
\usepackage{verbatim} 
\usepackage{multirow,bigdelim} 
\usepackage{color}
\usepackage{vmargin}
\setmarginsrb{1.2cm}{1cm}{1.2cm}{1cm}{1cm}{1cm}{1cm}{1cm}

\newcommand{\Mgas}{{M_{\rm gas}}}

  \makeatletter
    \renewcommand{\paragraph}{\@startsection{paragraph}{4}{\z@}%
      {-3.25ex\@plus -1ex \@minus -.2ex}%
      {1.5ex \@plus .2ex}%
      {\normalfont\small\centering}}
     
    \renewcommand{\subparagraph}{\@startsection{subparagraph}{5}{\z@}%
      {-3.25ex\@plus -1ex \@minus -.2ex}%
      {1.5ex \@plus .2ex}%
      {\normalfont\small\centering}}
    \makeatother

\newcommand{\kms}{{\rm km\,s^{-1}}}
\newcommand{\hMpc}{{ $h^{-1}$~Mpc}}

\title[Galaxies in the gap]{Properties of Galaxies in the disc central surface brightness gap}
\author[Sorce et al.]
{Jenny G. Sorce\thanks{E-mail: \texttt{jsorce@aip.de}},
Peter Creasey,
Noam I. Libeskind\\
Leibniz-Institut f\"{u}r Astrophysik, 14482 Potsdam, Germany\\
}

\begin{document}

\date{}

\pagerange{\pageref{firstpage}--\pageref{lastpage}} \pubyear{2015}

\maketitle

\label{firstpage}

\begin{abstract}
Intermediate surface brightness (ISB) galaxies are less numerous than their counterparts at high and low surface brightness (HSB and LSB). Investigating ISB characteristics from a sample from the S$^4$G survey, complete down to M$_B$=-16, we find that they have intermediate stellar, gas and baryonic masses and on average as much gas as stars. They lie on the (baryonic) Tully-Fisher relation between HSBs and LSBs, although they present a higher scatter than the latter. Their stellar to baryonic mass ratios have intermediate values unlike their condensed baryonic fractions. By comparing their environments, as classified by the eigenvalues of the velocity shear tensor of local constrained simulations, ISBs have a 5-10\% probability higher (smaller) to be in sheets (filaments) with respect to HSBs and LSBs. Additionally, for galaxies in filaments (with close neighbors), the mass and $\mu_0$ are correlated at 2.5 (2) $\sigma$ more than for those in sheets. ISBs live in regions where the divergence of the velocity field is smaller than where HSBs and LSBs live, a result at more than 50\% significance. ISBs may exist as an unstable transition state between LSBs and HSBs, the low flow activity environment maximally encouraging their formation. Interaction events altering the central baryon fraction could happen at a lower rate in these less dense environment, whilst in the higher density environments the LSBs are primarily satellite galaxies, whose accretion is sufficiently constrained that it fails to promote them to HSBs.
\end{abstract}
\begin{keywords}
galaxies: photometry; infrared: galaxies; galaxies: structure; galaxies: fundamental parameters 
\end{keywords}


\section{Introduction}
 
The observed distribution of galaxy surface brightness profiles as well as that of other galaxy parameters are fundamental for understanding galaxy formation. The evolution from a Gaussian distribution of density fluctuations to a distribution of galaxies that can be classified into distinct categories provides strong constraints on the evolutionary process \citep{1985ApJ...295...73K,1986ApJ...303...39D,1998ARA&A..36..435M}. As an example, the distribution of disc central surface brightnesses of galaxies ($\mu_0$) exhibits a deficiency at intermediate values \citep{1997ApJ...484..145T,2009MNRAS.393..628M,2009MNRAS.394.2022M,2013MNRAS.433..751S}. In this paper we focus on this intermediate population to analyse whether it is atypical in other aspects.

De Vaucouleurs first studied surface brightness profiles in 1948. He was followed by \citet{1959Obs....79...54S} and \citet{1970ApJ...160..811F}.  However, the $\mu_0$ bimodal distribution was first highlighted only several years later by \citet{1997ApJ...484..145T}.  The study results showed that the $\mu_0$ distribution in Ursa Major was discontinuous: there was a lack of galaxies of intermediate surface brightness or, from another perspective, an excess of low and high surface brightness (L/HSB) galaxies. Two stable modes for galaxy formation were then suggested. A mode where dark matter dominates at all radii giving rise to LSB galaxies and a central baryonic matter domination giving birth to HSB galaxies. It was suggested that the cause of ISBs being less abundant than their HSB or LSB counterparts could be the result of galaxies avoiding the situation where baryonic and dark matters are co-dominant in the centre, ISB galaxies being a transition state between LSB and HSB. 
However, concerns were expressed of large possible errors in fitting galaxy discs  due to shallow K'-band observations which could lead to premature truncation of discs \citep{1997ApJ...484..145T}. Bulges could be partially included in fits leading to a bias in $\mu_0$. \citet{2000MNRAS.311..668B} also argued that the bimodality could also be an artefact due to incorrect inclination-corrections applied to the $\mu_0$ values or that it could result from small number statistics. A larger number of galaxies in the Virgo cluster was studied by \citet{2009MNRAS.394.2022M} partitioning bulgeless and with bulge galaxies and using multi-component fits. They concluded with the existence a dichotomy for $\mu_0$ too. Still, with only two clusters presenting this dichotomy, they were hesitant to claim that bimodality is not a cluster effect. \citet{2013MNRAS.433..751S} studied the distribution of disc central surface brightnesses for a complete sub-sample of galaxies in the Spitzer Survey of Stellar Structure in Galaxies \citep[S$^4$G,][]{2010PASP..122.1397S}. Their sample extended to 20 Mpc and included all the galaxies brighter than M$_B$=-16. They concluded that indeed galaxies seem to avoid prolonged occupation of a state where dark and baryonic matter are co-dominant in the centre.

ISB galaxies have been found repeatedly to be less abundant than their HSB and LSB counterparts and hypotheses about their low number and the distribution of matter within their core have been made \citep[e.g.][]{1999ASPC..182..383V,1999ASPC..170...92V}. They have yet to be investigated in terms of their different characteristics such as masses and locations in the cosmic web environment with respect to HSB and LSB galaxies. Recently, a variety of parameters derived for galaxies from S$^4$G has been released \citep{2015ApJS..219....3M,2015ApJS..219....4S}. 

In this paper, we use the set of parameters obtained for a selected complete sample of galaxies in S$^4$G. In section 2, the selected sample is presented and the dichotomy in the disc central surface brightness classification is verified using the 2-dimensional structural surface brightness decompositions of 3.6~$\micron$ images obtained with GALFIT3.0 \citep{2010AJ....139.2097P,2015ApJS..219....4S}. In section 3, galaxies in the gap or intermediate disc central surface brightness galaxies are the object of focus. Parameters such as different masses (baryonic, stellar, etc) are derived for both these galaxies and HSB / LSB galaxies. These derived parameters are compared with those of HSB and LSB galaxies. In section 4, the environment of ISBs is analyzed using the cosmic web definition applied to local constrained simulations resembling the local universe within 30 \hMpc\ down to a few megaparsecs \citep[typically 2-3 \hMpc,][]{2014MNRAS.437.3586S} and compared to that of HSBs and LSBs. We conclude proposing an explanation for the observed distribution of ISB galaxies.

\section{Disc Central Surface Brightness Distribution}

\subsection{Galaxy Sample}
The $S^4G$ survey is a volume-, magnitude-, and diameter-limited ($d<40$~Mpc, $|b|>30^\circ$, $m_{Bcorr}<15.5$ and $D_{25}>1'$) survey of over 2,350 galaxies observed at 3.6 and 4.5\micron\  (channels 1 and 2 respectively) with the IRAC instrument \citep{2004ApJS..154...10F} aboard the Spitzer Space Telescope \citep{2004ApJS..154....1W}. It is a very large extremely deep, representative and homogeneous sample of nearby galaxies containing all morphological types.\\

\citet{2015ApJS..219....3M} derived the 1D-photometric parameters of all the galaxies in the 3.6 and 4.5\micron\ band data. \citet{2015ApJS..219....4S} use only the 3.6\micron\ band data for the 2D decomposition. The preference for the [3.6] band over the [4.5] band is based on the knowledge that 4.5\micron\ fluxes have a higher contribution from hot dust than fluxes at 3.6\micron\ \citep{2012AJ....144..133S,2012ApJ...744...17M}. Every galaxy of type later than $S0^-$ (-3, for the numerical value), the latter included\footnote{Note that tests showed that including or excluding $S0^-$ galaxies does not affect the conclusions.}, up to a distance of roughly 20 Mpc obtained with the apparent and absolute magnitudes given by \citet{2015ApJS..219....3M}, is selected. This subsample goes down to an absolute magnitude limit of -16 in the B band in the Vega magnitude system. The faint magnitude limit prevents the loss of LSB galaxies \citep[][]{2008MNRAS.391..986Z} from the volume surveyed and guarantees the presence of ISB galaxies.\\

\subsection{Disc Central Surface Brightness}
\begin{figure}
\vspace{-1.4cm}
\hspace{-0.2cm}\includegraphics[width=0.55 \textwidth]{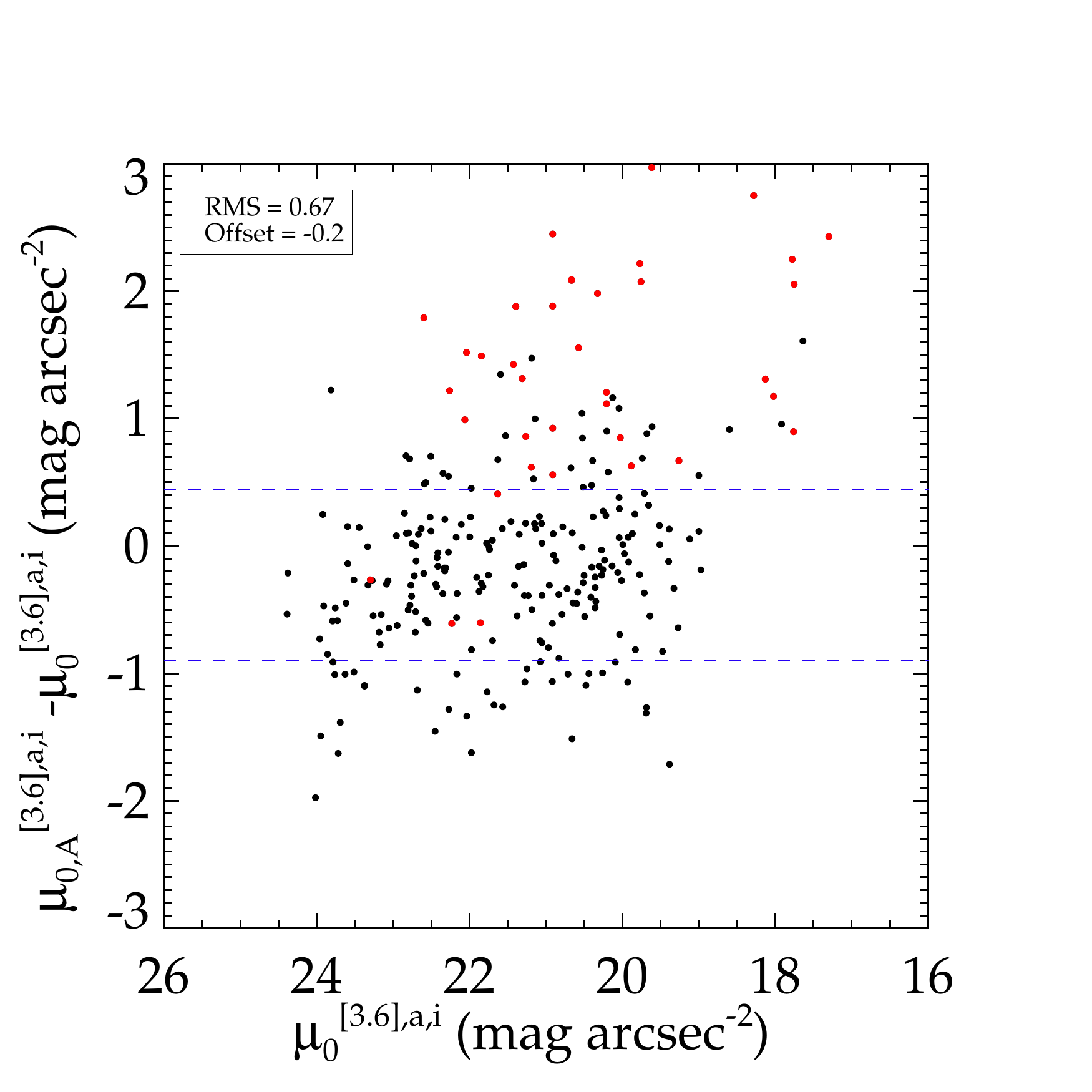}
\vspace{-0.5cm}
\caption{Residual between Spitzer-adapted Archangel  $\mu_0$ obtained with 1D disc-only fit and S$^4$G $\mu_0$ obtained with 2D decomposition versus $\mu_0$ obtained with 2D decomposition in mag arcsec$^{-2}$. Red dots stand for edge-on galaxies. Offset (red dotted line) and standard deviation (blue dashed lines) about it are computed without considering edge-on galaxies.}
\label{fig:comp}
\end{figure}

\citet{2015ApJS..219....4S} supply us with $\mu_0$ de-projected for inclination. Inclination effects can be quite confusing at optical band because the path length of observed surface brightnesses varying with the line-of-sight, and extinction, work in opposite way on measured $\mu_0$ values.  Since at 3.6 microns obscuration can be assumed to be negligible and therefore only the geometric effect of the inclination on surface brightnesses needs to be taken into account, $\mu_0$ values given by \citet{2015ApJS..219....4S} for the disc component are retained without further correction.\\
In Figure~\ref{fig:comp} we compare the residuals between the $\mu_0$ values of \citet{2013MNRAS.433..751S} obtained with the Spitzer-adapted Archangel software \citep{2012AJ....144..133S} and those of \citet{2015ApJS..219....4S}. \citet{2013MNRAS.433..751S} used 1D disc-only fits to obtained $\mu_0$ and they corrected them with the following formula for inclination:
\begin{equation}
\mu_0^{[3.6],a,i} = \mu_0^{[3.6],a}-2.5C^{[3.6]} log (\frac{b}{a})
\label{eq:inccorr}
\end{equation}
with $C^{[3.6]}$=1 assumed for a transparent system. By extrapolation, this formula matches equation 4 of \citet{2015ApJS..219....4S}. $\mu_0^{[3.6],a,i}$ is the disc central surface brightness corrected for both aperture and inclination, and $\mu_0^{[3.6],a}$ is the disc central surface brightness corrected for aperture corrections. While \citet{2015ApJS..219....4S}  use b/a ratios of the S$^4$G pipeline to derive inclination, \citet{2013MNRAS.433..751S} derived these later from b/a ratios of the HyperLeda database \citep{2003A&A...412...45P}. The aperture correction is required for large extended source, like galaxy, photometry for the IRAC instrument normalized to a finite aperture  \citep[12",][and IRAC Instrument Handbook]{2005PASP..117..978R}. The same recommended correction\footnote{http://irsa.ipac.caltech.edu/data/SPITZER/docs/irac/iracinstrumenthandbook/} for fluxes was applied by the S$^4$G team (e.g. M\~unoz-Mateos et al. 2015).

In Figure \ref{fig:comp}, the offset of residuals between the two measurements and the standard deviation of data points about it are computed removing all galaxies which are assumed to be edge-on by \citet{2015ApJS..219....4S}. The latter defined a galaxy to be edge-on if its axial ratio, b/a, is less or equal to 0.2. $\mu_0$ of edge-on galaxies are difficult to characterize because integrating along the line of sight may hide the effects of sub-structures like bars and spiral arms \citep[e.g.][]{2010MNRAS.401..559M}. No precise method exists for correcting such effects and the coefficient $C^{[3.6]}$ in equation \ref{eq:inccorr} itself may vary with galactic radius. In the absence of a superior method to correct for inclination, every edge-on galaxy is then removed from the offset and standard deviation comparison. A standard deviation of 0.7 mag arcsec$^{-2}$ and an offset of -0.2 mag arcsec$^{-2}$ are found. The sign of the offset is the same as for that derived when comparing apparent magnitudes obtained by the S$^4$G software and the Spitzer-adapted Archangel software \citep[-0.02 mag,][]{2014MNRAS.444..527S}. There is a tendency for S$^4$G $\mu_0$ values to be higher for LSB galaxies and smaller for HSB galaxies, however this primarily only occurs when the edge-on galaxies are included. 
Causes for this, apart from the different type of fits (1D versus 2D) include differences in inclinations (b/a ratios computed in the optical vs b/a ratios derived at 3.6 microns) and in masking. Distributed equally, the 0.7~mag~arcsec$^{-2}$ scatter gives an uncertainty about $\pm 0.47$~mag~arcsec$^{-2}$ for each source. Regardless, it is reassuring that both disc central surface brightnesses are in agreement with these two alternative computations, especially a 1D disc-only fit and a 2D decomposition. The scatter is in relative agreement with values given by \citet{1996A&AS..118..557D}, where disc-only fits were showed to result in unbiased disc parameters relative to 2D fit decomposition parameters within 0.5 mag arcsec$^{-2}$. The agreement is appreciable especially considering the different inclinations and masks. Since the background (distant galaxies and zodiacal light) uncertainties lead to a small magnitude uncertainty even for IRAC ch1 \citep{2012AJ....144..133S}, an uncertainty budget of 0.5 mag arcsec$^{-2}$ for $\mu_0^{[3.6]}$ measurements is retained. The bin size of 0.5 mag arcsec$^{-2}$ for histograms of $\mu_0^{[3.6]}$  distributions chosen by \citet{2013MNRAS.433..751S} is preserved.

\subsection{Bimodality}
\begin{figure}
\vspace{-1.4cm}
\hspace{-0.3cm}\includegraphics[width=0.55 \textwidth]{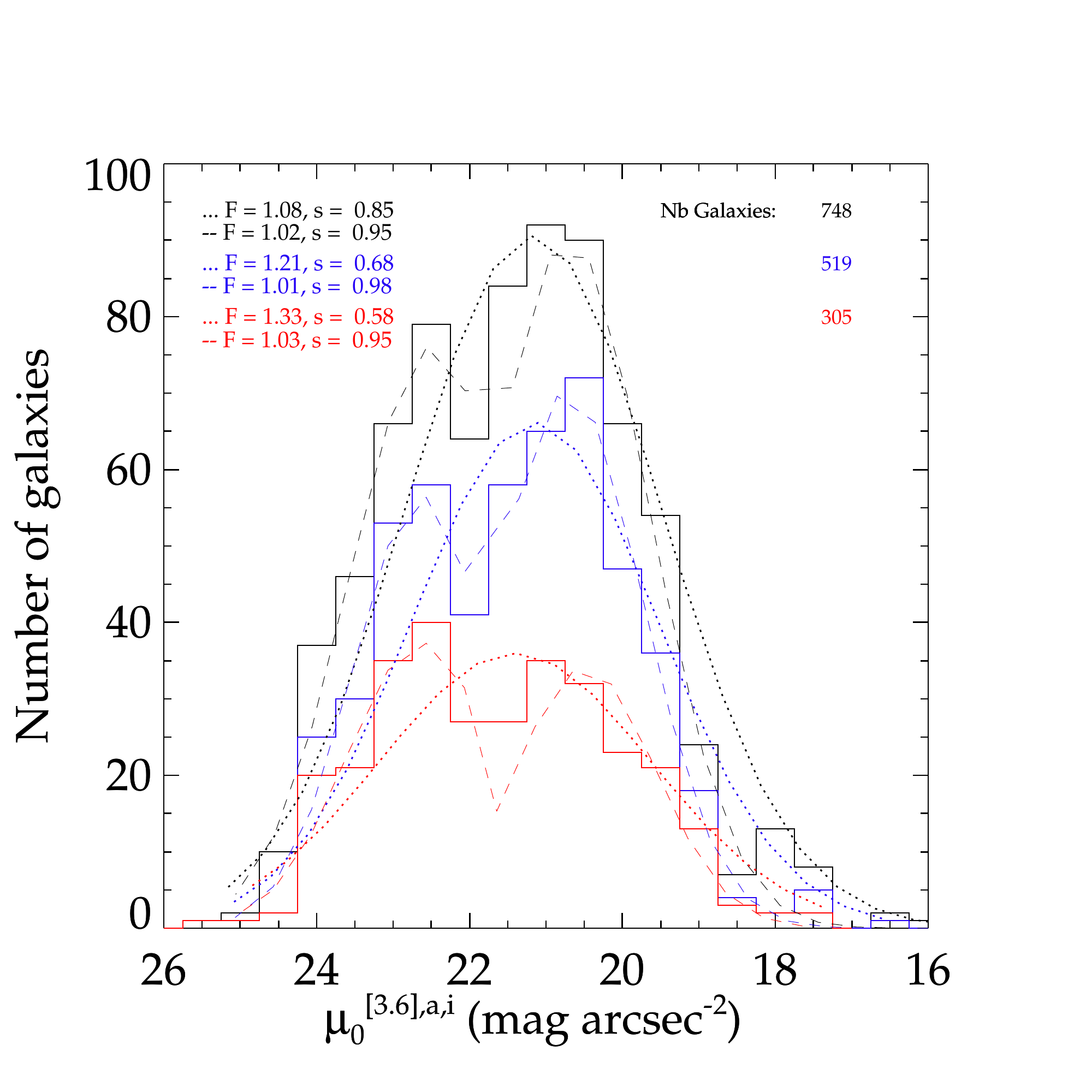}\\
\vspace{-1.cm}

\hspace{-1.7cm}\includegraphics[width=0.59 \textwidth]{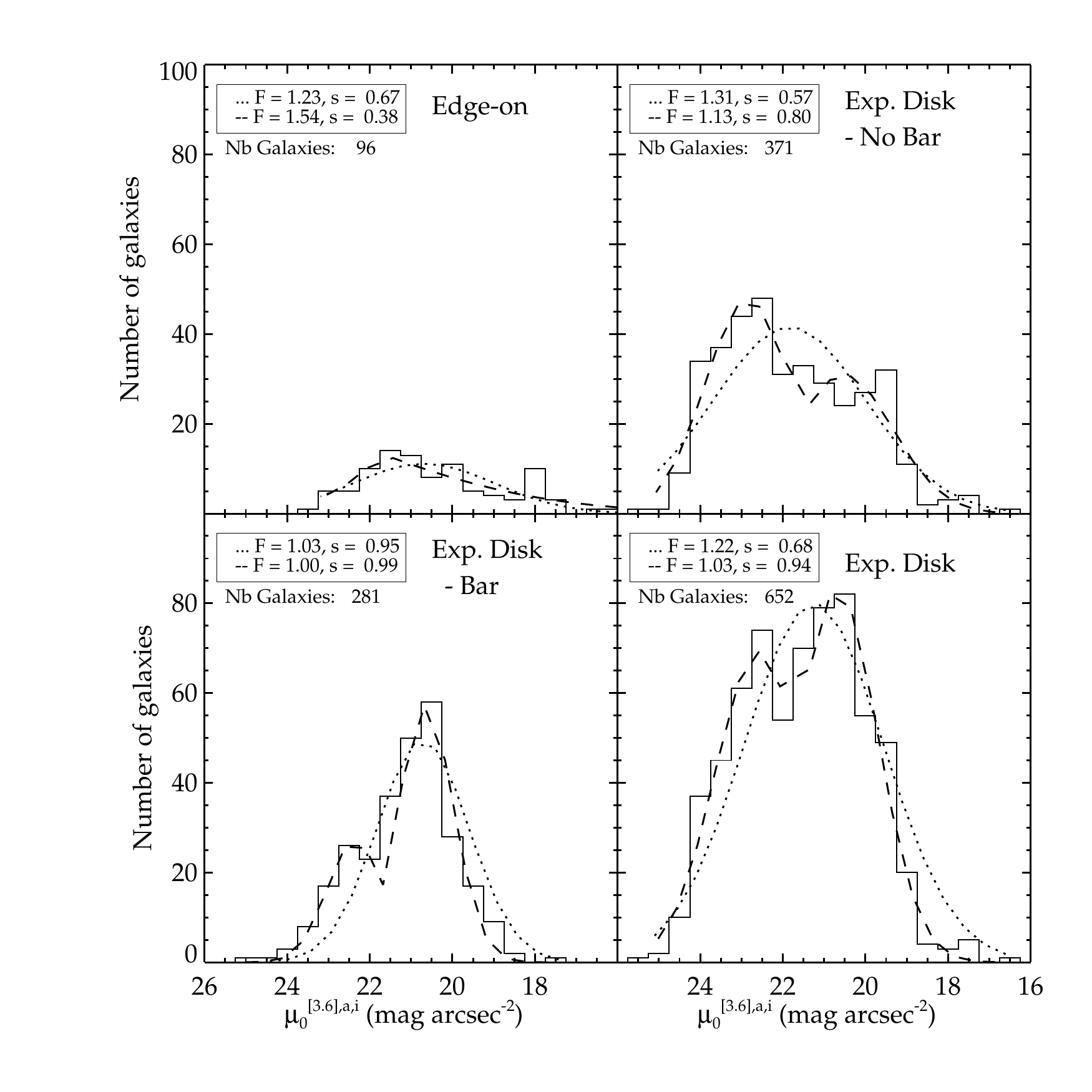}
\vspace{-1cm}

\caption{Top: Distribution of disc central surface brightnesses for the entire (highest black solid line histogram) S$^4$G selected galaxy sample (at less than 20~Mpc and with a type later than S0$^-$), for galaxies less inclined that 71$^\circ$ (intermediate blue solid line histogram) and for galaxies of intermediate inclinations (between 45$^\circ$ and 71$^\circ$, smallest red solid line histogram). In the three cases, F-tests (F) show that the variance of the galaxy sample is closer to that of the double Gaussian least square fit (dashed lines) than to that of the single Gaussian least square fit (dotted lines). Bottom: the same as the top panel but for the entire set of edge-on discs  (top left), for all the exponential discs  without (top right) and with (bottom left) bars and all the exponential discs  together (bottom right). The only $\mu_0$ distribution which reflects a higher significance level (s) in having a variance closer to that of its single Gaussian fit (dotted line) rather than to that of its double Gaussian fit (dashed line) is that of edge-on discs. However, the quality of $\mu_0$ measurements for edge-on discs  is difficult to assess as structures can be hidden.}
\label{fig:bimo}
\end{figure}

The distribution of $\mu_0$ is displayed by a black solid line on the top panel of Figure \ref{fig:bimo} where two peaks are visible. The choice of bin sizes, although motivated by the conservative estimate for the error on $\mu_0$ of 0.5 mag arcsec$^{-2}$, might strongly influence the appearance of the histogram: a small bin size could give rise to a multitude of spurious peaks while a large bin size could mask the bimodality. Instead of assessing the nature of the distribution only with the visual aspect of histograms, as the disc central surface brightness is a priori a continuous variable characterized by its probability distribution function, we consider single and double normal distributions. Both single and double Gaussians are then fitted to the histogram. The final aim is to determine whether the $\mu_0$ distribution has two distinct peaks or not, i.e. whether the distribution resembles more the double or the single Gaussian least square fits in terms of moments (e.g. variance). To quantify this and thereby the nature of the distribution (i.e. uni/bi-modal), F-tests \citep[based on the variance of the data to variance of the fit ratio, denoted `F', see e.g.][]{1989Snedecor} are applied. To derive the significance the degrees of freedom are the numbers of bins minus 1. If the ratio F is either too large or too small, the hypothesis, that the variance of the dataset is equal to the variance of the fit, can be rejected at a high significance level (denoted `s'); the ``fit'' being either a double or single Gaussian (bimodal or unimodal respectively). Figure \ref{fig:bimo} reveals that the variances of both fits are similar to the variance of the observational sample although there is a slightly better agreement between the variances of the dataset and of the double Gaussian fit (black dashed line) than with that of the simple fit (black dotted line) at the 10\%\ level. Still, the position of the potential gap is at $\sim$~22~$\pm$~0.5~mag~arcsec$^{-2}$ similar to that of \citet{2013MNRAS.433..751S} including the uncertainty and the shift between the two different measurements of $\mu_0$. Additionally as explained in section 2.2, highly inclined galaxies constitute difficult cases. Thus, the distribution, omitting galaxies more inclined that 71$^\circ$ is also plotted in blue solid line\footnote{We motivate our choice by those of \citet{2000MNRAS.311..668B}, $69^\circ$ and \citet{2013MNRAS.433..751S}, $73^\circ$. Regardless, because of the uncertainties on inclinations ($\sim$ $4-5^\circ$), choosing 69, 71 or 73$^\circ$ does not change the conclusions.}. Inclination values given by \citet{2015ApJS..219....4S} are used. Once the highly inclined galaxies are omitted, the bimodal distribution is reinforced since the agreement between the variances of the dataset and the double fit is increased to 98\%\ while that with the single fit is decreased some more (down to 68\%). Later on, masses will be computed to study more closely ISB galaxies, via the utilization of HI fluxes. Important relations, such as the baryonic Tully-Fisher correlation \citep{2000ApJ...533L..99M}, require rotation rate of galaxies which need to be de-projected for inclination. To limit errors in rotation rates due to de-projection, galaxies of inclination $<45^\circ$ are often discarded \citep[e.g.][]{2009AJ....138.1938C,2011MNRAS.414.2005C}. Accordingly, the red solid line represent the $\mu_0$ distribution for galaxies of intermediate inclinations (between 45 and 71$^\circ$). Hence, there is a high enough confidence in both $\mu_0$ and rotation rate measurements. The bimodality is still present at more than the 95\% significance level while the agreement with the single Gaussian distribution recesses to the 58\% significance level. An additional sanity check reveals that for each double Gaussian the residual between the two peaks is larger than twice their almost equal standard deviations.

\begin{figure}
\hspace{-0.2cm}\includegraphics[width=0.55 \textwidth]{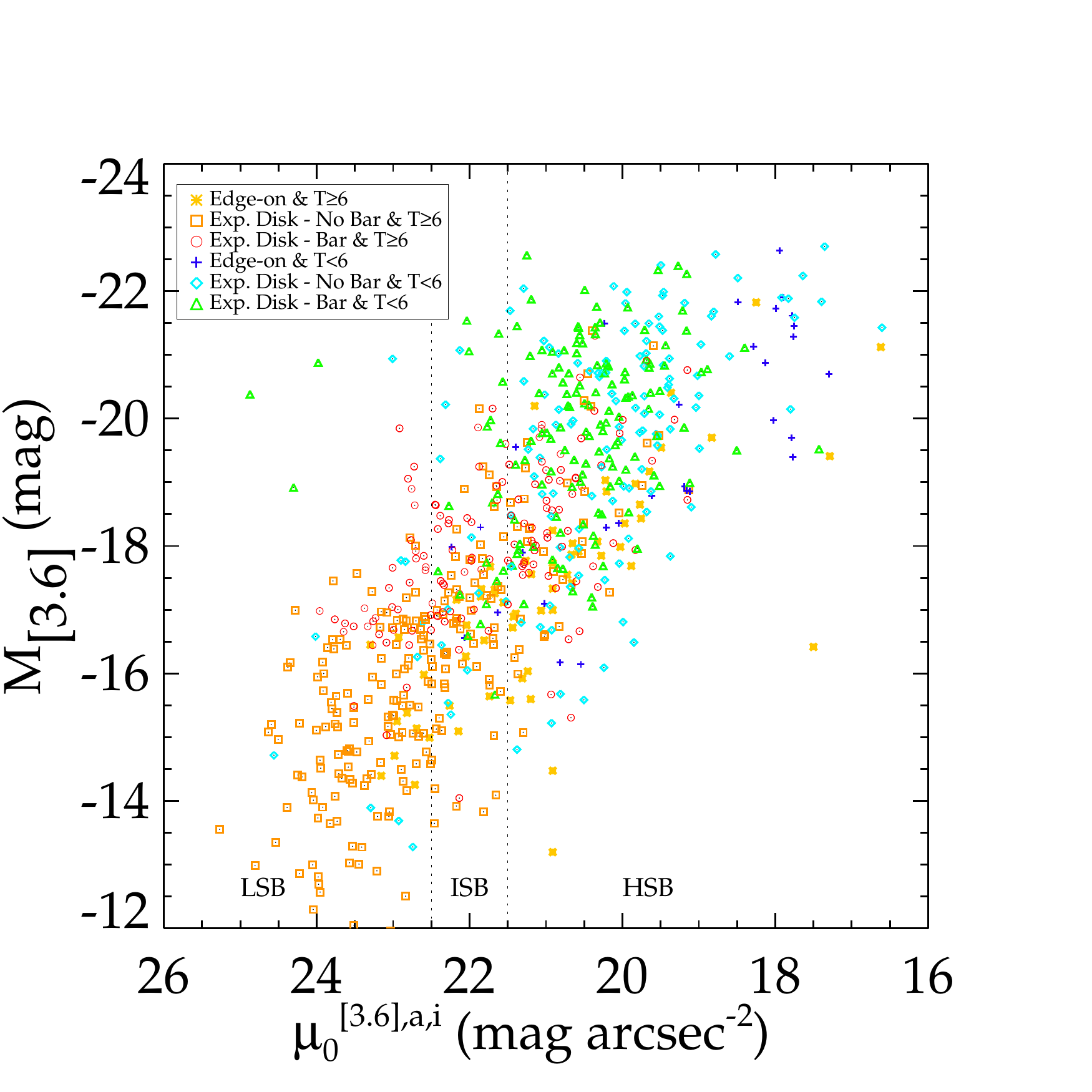}
\caption{Absolute magnitude at 3.6 micron versus disc central surface brightness in the same band. Galaxies are separated by morphological type (earlier, cold colours, and later, warm colours, than 6 - Scd) and by structures (edge-on, yellow circles and blue triangles, exponential discs  with, red stars and green crosses, and without, orange squares and light blue diamonds, bars). LSB, ISB and HSB regions can be defined. Galaxies off the relation have $\mu_0$ measurements mostly of less quality according to S$^4$G criteria. They are often highly inclined galaxies.}
\label{fig:distri}
\end{figure}

\begin{figure}
\hspace{-0.2cm}\includegraphics[width=0.55 \textwidth]{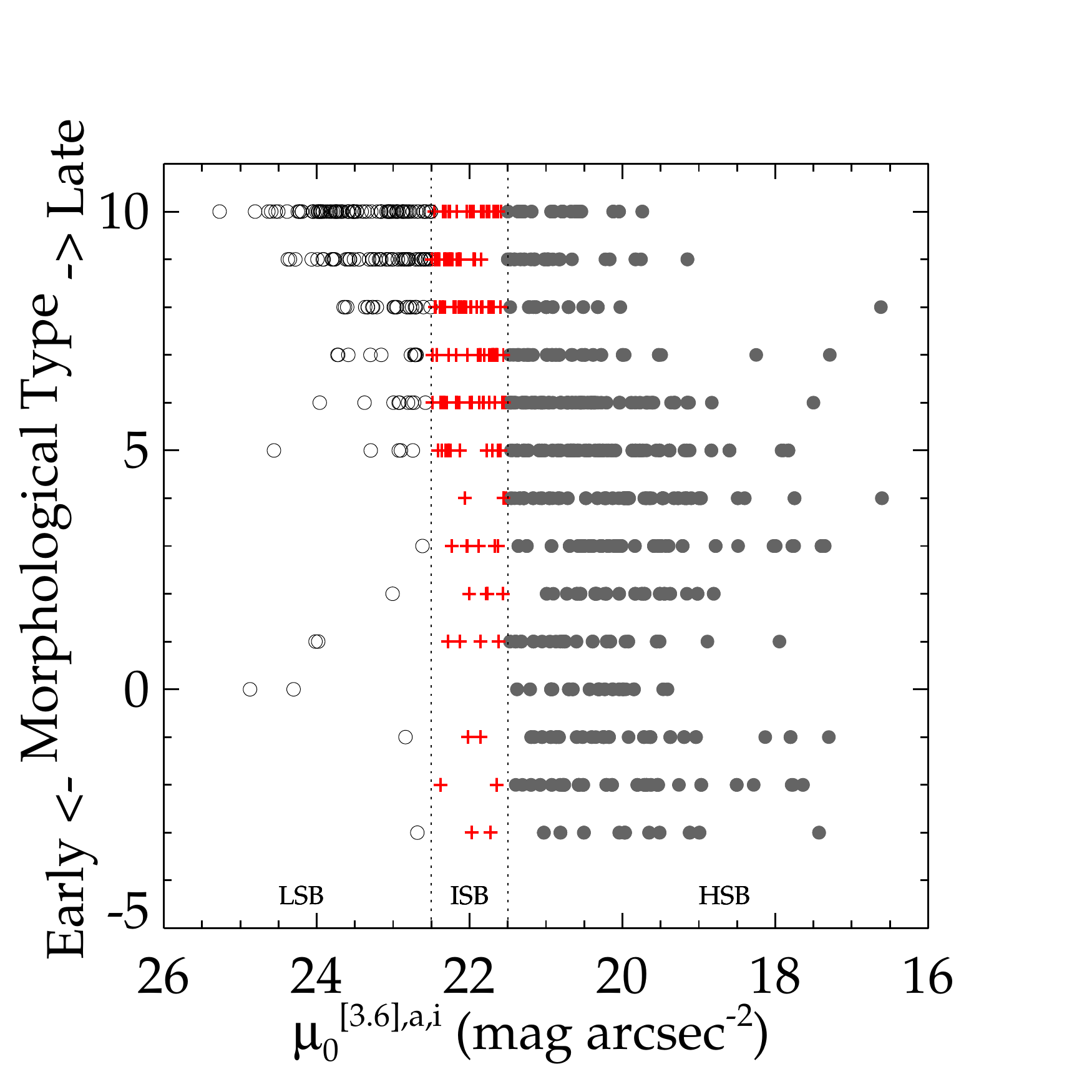}
\caption{Disc central surface brightness $\mu_0$ versus numerical morphological type. Open circles stand for LSBs while grey filled circles represent HSBs. Red crosses are ISBs (galaxies in the gap). ISBs are in majority of type 6 (Scd) or later.}
\label{fig:type}
\end{figure}

\citet{2010ApJS..190..147B,2015ApJS..217...32B} categorized galaxies in sets of edge-on, exponential disc with and without bars. The bottom panel of Figure \ref{fig:bimo} represents $\mu_0$ distribution for these different categories: edge-on discs  (top left quadrant), exponential discs without bars (top right quadrant) and with bars (bottom left quadrant) and the entire set of exponential discs (last quadrant). As already observed by \citet{2013MNRAS.433..751S}, disc central surface brightness of edge-on discs are difficult to assess. It was already hinted at by the discrepancy between the $\mu_0$ values measured by the adapted Archangel software compared with the S$^4$G pipeline for these galaxies. As a result, their $\mu_0$ distribution is in better agreement with the single Gaussian (dotted line) than the double Gaussian (dashed line) fits. In contrast, however, exponential discs  without bars, with bars or irrespective of bars are better represented by a double Gaussian than a single one. An additional sanity check reveals that the residual between the two peaks is larger than twice their standard deviation.  As expected, $\mu_0$ histograms show that exponential discs  with bars have in general brighter disc central surface brightness than those without bars. Bars trigger star formation and thus enhance luminosity in galaxy centres \citep[e.g.][]{2011MNRAS.416.2182E,2011ApJ...734..122M}. 

Figure \ref{fig:distri} shows absolute magnitude versus $\mu_0$ distinguishing edge-on, exponential disc with and without bar galaxies. Earlier and later types (T $<6$ and $\ge 6$ respectively, T=6 corresponds to Scd galaxies) are separated. The Figure allows the discrimination of regions for LSB, ISB and HSB galaxies. The LSB region is primarily populated by late type galaxies while the HSB region contains a large number of earlier types. Galaxies offset from the relation between absolute magnitude at 3.6 micron and $\mu_0$ measured in the same band, have $\mu_0$ measurements of less quality according to S$^4$G criteria \citep{2015ApJS..219....4S} and are often highly inclined galaxies. Galaxies in the $\mu_0$ distribution gap can either be galaxies with or without bars but are in general of intermediate or later type (around 6 and later) as shown also on Figure \ref{fig:type}. In the next section, characteristics, in particular masses, of ISB galaxies are derived and compared with those of LSB and HSB galaxies.

\section{Stellar, Gas, Baryonic and Total Masses}
  
  \begin{figure*}
\vspace{-2cm}
\includegraphics[width=0.54 \textwidth]{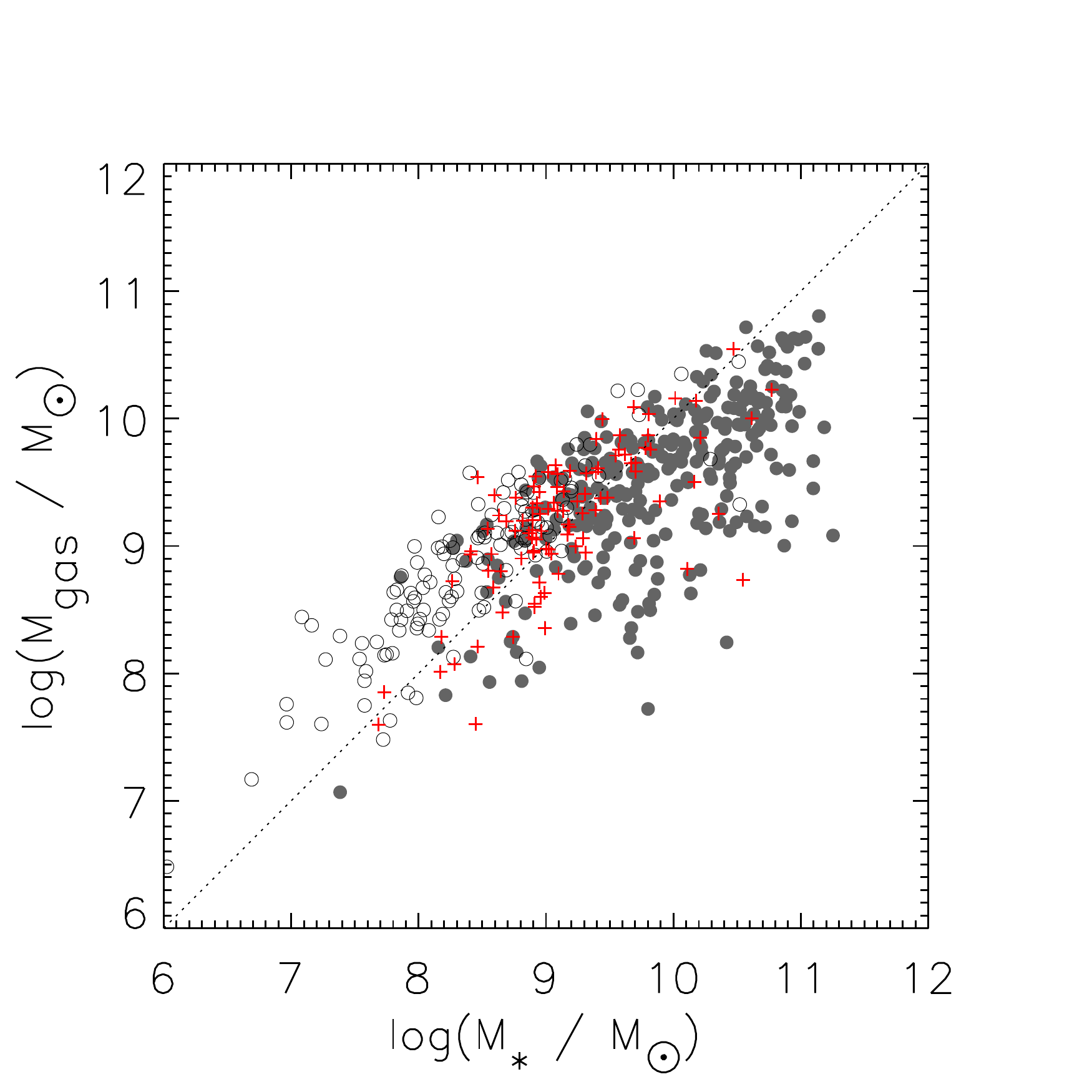}
\hspace{-1.65cm}\includegraphics[width=0.54 \textwidth]{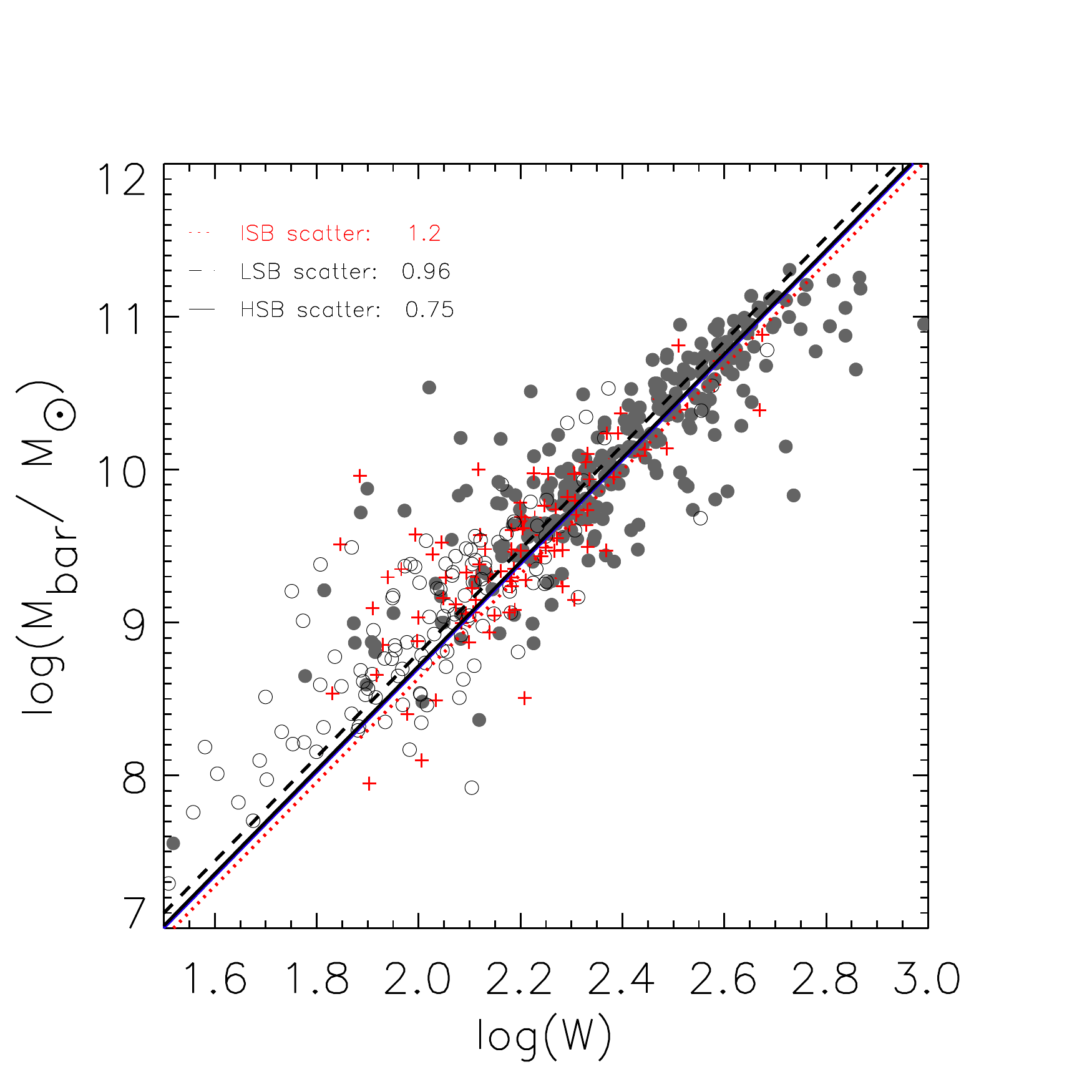}\\
\vspace{-0.25cm}
\caption{Relation between galaxy stellar and gas masses (left) and galaxy baryonic mass and linewidth adjusted to twice the rotation rate (right). Open black circles stand for LSBs, filled grey circles represent HSBs. ISBs are displayed as red crosses. In the left panel, the dotted line is the 1:1 relation to enhance the clear separation between LSBs, above the line and HSBs, below the line while ISBs are on the line. The right panel is the correlation between baryonic mass (defined as the sum of stellar and gas masses) and linewidth (twice the rotation rate) of galaxies or baryonic Tully-Fisher relation. The solid line stand for the relation derived for HSB galaxies assuming a 3.4 slope. The Milky Way fits on that line. Dashed and dotted lines are similar fits but to LSBs and ISBs respectively. While keeping the slope fixed, zero points have to be slightly modified to minimize scatters.}
\label{fig:relation}
\end{figure*}

In this section we derive stellar, gas and baryonic masses of galaxies to investigate the nature of ISB galaxies with respect to that of HSBs and LSBs. To derive gas masses, HI profiles of galaxies are needed. We use HI fluxes from the Extragalactic Distance Database\footnote{http://edd.ifa.hawaii.edu} \citep[][]{2009AJ....138..323T}. M$_*$ are given by \citet{2015ApJS..219....3M}. We refer to the baryonic mass as the sum of stellar, ($M_*$), and gas masses ($\Mgas$) calculated from the relations
 \begin{eqnarray}
 \Mgas &=& 1.4 \times [(3.7-0.8T+0.043T^2) + 1]\times \nonumber \\
&& (2.36 \times 10^5 \times D_L^2 \times F_{HI}) \frac{M_\odot}{10^5} \nonumber \\
 M_*&=&10^{5.65}\ F_{[3.6]}^{2.85}\ F_{[4.5]}^{-1.85}\ (\frac{D_L}{0.05})^2\ M_\odot
 \label{eq:mbaryonic}
 \end{eqnarray}
from \citet{1997ApJ...481..689M,2014AJ....147..134Z}  and from \citet{2012AJ....143..139E}. $T$ is the numerical morphological type. In $\Mgas$, the 1.4 factor accounts for helium and metals, the expression with $T$ corrects for the molecular hydrogen and the rest gives the HI mass. In a few cases the formula for hydrogen results in an unphysical negative (but small) value, in which case it is set to zero. $F_X$ are fluxes (integrated within the HI profile in units of Jy $\kms$ or in [3.6] and [4.5] Spitzer bands in Jy). $D_L$ is the luminosity distance. 

To pursue the study, the left panel of Figure \ref{fig:relation} displays the relations between galaxy gas and stellar masses. Galaxies classified as LSBs, HSBs and ISBs are displayed with different symbols or colours: LSBs are represented by open black circles while HSBs are plotted as filled grey circles ; The red crosses stand for the ISB galaxies. The dotted line is the 1:1 relation to enhance the clear separation between LSBs, HSBs and ISBs: HSB galaxies have higher stellar masses (below the line) while LSBs have a higher content of gas (above the line); The ISBs contain approximately as much gas as luminous material (on the line). 

\begin{figure}
\hspace{-1cm}\includegraphics[width=0.55 \textwidth]{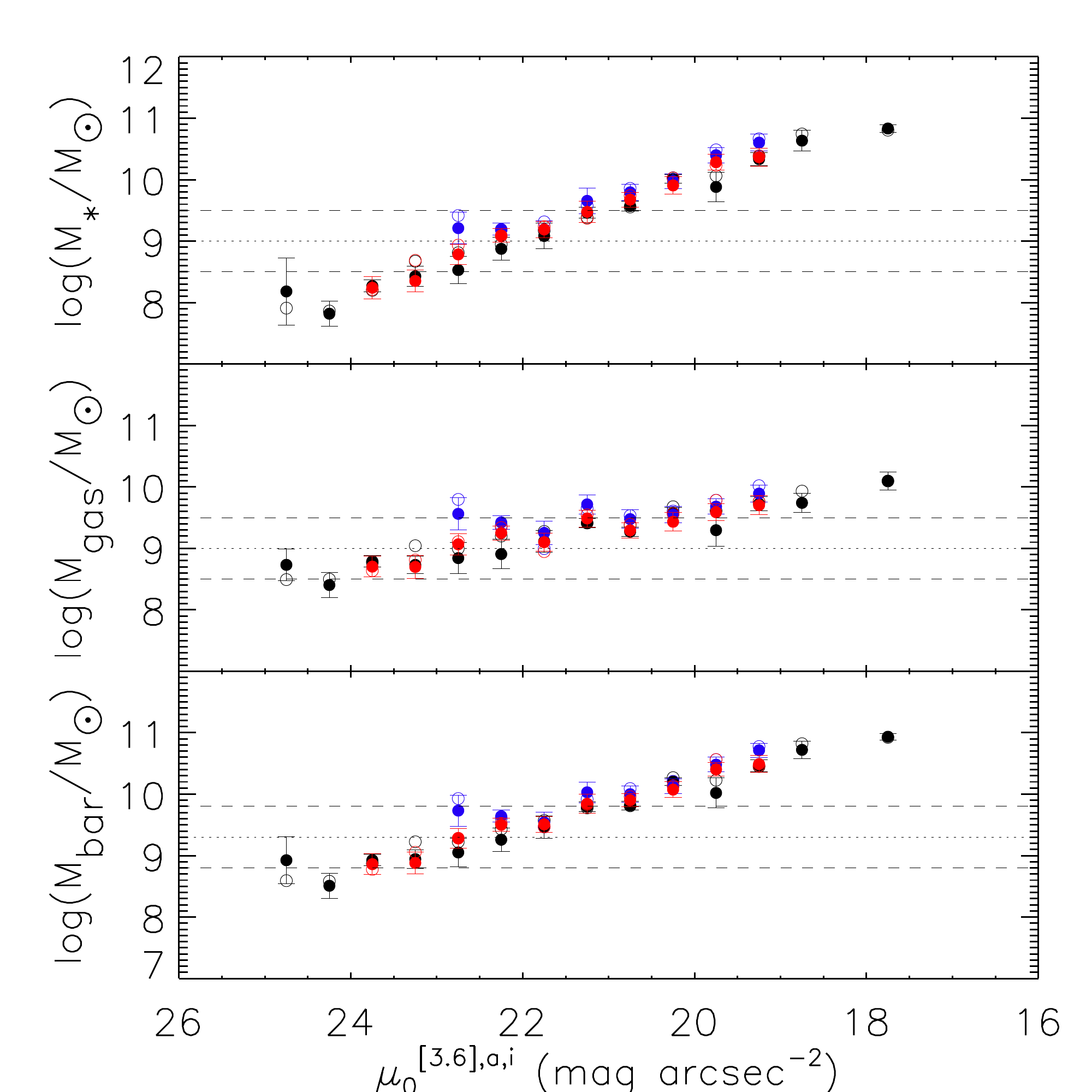}
\caption{Mean (filled circles) and median (open circles) of stellar (top), gas (middle) and baryonic (bottom) masses as a function of disc central surface brightness. Bars represent the standard error of the means. A bin size of $0.5\, \rm  mag\, arcsec^{-2}$ is used to compute mean and median values. Black points include the entire sample (spanning over the largest range of $\mu_0$), the red color stands for the sample excluding galaxies with extreme inclinations (below 45$^\circ$ or above $71^\circ$, spanning over an intermediate range of $\mu_0$) and the blue one additionally excludes very late (later than Sdm) or early (earlier than S0) type galaxies (spanning over the smallest range of $\mu_0$). A mean (median) is marked only if at least 5 galaxies contribute to the computation of the value. In the upper two panels, the dotted lines mark the average value of gas and stellar masses required for a galaxy to have them contribute equally to the baryonic mass (dotted line in the lowest panel), i.e. the dotted lines in the first two panels indicate M$_*$=M$_{gas}$=0.5 M$_{bar}$ and the dotted line in the bottom panel marks M$_{bar}$=M$_*$+M$_{gas}$=2M$_*$=2M$_{gas}$. Dashed lines show the limit of such a possible equal contribution on average. Only galaxies in the gap have on average as much gas as luminous material.}
\label{fig:mass}
\end{figure}

This is confirmed, on average, by Figure~\ref{fig:mass}, where the mean and median stellar, gas and baryonic masses of the galaxies are given as a function of disc central surface brightness. A bin size of 0.5~mag~arcsec$^{-2}$ is used to compute each value. The black colour corresponds to the entire sample, the red to that of non-extreme inclination galaxies ($45$-$71^\circ$) and the blue one additionally excludes those of extreme morphological type (galaxies between Sb, T=3 included and Sdm, T = 8 included are kept). On average, LSB galaxies have more gas than luminous material while it is the opposite for those of HSB. The ISB sample has stellar and gas masses which contribute equally to their baryonic masses. On average, solely galaxies in the gap can have such a repartition of baryonic mass. The ISB galaxies have intermediate stellar, gas and baryonic masses with respect to LSBs and HSBs and they contain as much gas as stars.

We next focus our interest on ISB position in the luminosity-rotation rate correlation \citep{1977A&A....54..661T}. Linewidth values, adjusted to twice projected rotation rate values, come from the `All Digital HI' catalogue \citep[e.g.][for a detailed description of the adjustment]{2011MNRAS.414.2005C} of the Extragalactic Distance Database \citep[EDD][]{2009AJ....138..323T}. As recommended, we keep linewidth values when errors are less than $20\,\kms$. 

Initially the total sample is kept. A simple linear fit, (although not optimal since galaxy types earlier than Sa-Sb are included), of the Tully-Fisher relation reveals a strong correlation with central surface brightness: the ISBs lie in between HSBs and LSBs on the fit. There are a few outliers (LSBs lying in the middle of HSBs or the opposite) probably due to the opposite sign effect of inclinations on the uncertainties of linewidth and $\mu_0$ estimates. Namely, while linewidths estimates are more accurate for highly inclined galaxies, $\mu_0$ values are more accurate for low inclination galaxies and vice-versa. Errors grow in opposite direction: typically a galaxy with a very accurate $\mu_0$ is thus likely to have a conversely low accuracy linewidth and be misplaced in the Tully-Fisher diagram.

The goal of this paper is not to derive an absolute relation (we refer the reader to \citealp{2013ApJ...765...94S,2014MNRAS.444..527S} for a thorough analysis) but to compare the relative scatter for the different surface brightnesses value galaxies. To facilitate comparisons with other relations absolute magnitudes are replaced by logarithms of stellar masses and a simple fit is computed again. Similar fits are also derived for the logarithms of the other (gas and baryonic) masses versus logarithms of linewidths. The higher scatter ($>$ 1.0~dex) is derived for the relation between gas mass and linewidth implying that gas masses are not very well correlated with rotation rate (half the linewidth). On the other hand, the relation between stellar mass and linewidths is found to be tight ($<$ 1.0~dex) conforming to our expectations for the Tully-Fisher relation, i.e. the luminous content is a good proxy for galaxy masses. Still the correlation is tighter (0.8 dex) when using the baryonic content as found by \citet[][at 3.6\micron ]{2014AJ....147..134Z}. This is the relation chosen to be shown on the right panel of Figure \ref{fig:relation}. 
This is the most judicious choice as we are in presence of a large range of galaxy masses, in particular the LSB population includes many gas-dominated systems that are not well fit by the classical Tully-Fisher relation, in that case the gas component is not negligible in comparison to the luminous stellar component.
Anomalous galaxies - in the sense that they are not classic disc dominated spirals - also deviate from the classical Tully-Fisher relation. The baryonic Tully-Fisher relation is able to reconcile disagreements which can be found between such different systems, namely they do lie on the same baryonic Tully-Fisher relation but have their own classic Tully-Fisher relation \citep[e.g for massive disc galaxies and for Hickson compact group galaxies,][]{2007MNRAS.381.1463N,2013MNRAS.432.3085T}. ISB galaxies tend to exhibit intermediate masses (gaseous, stellar or baryonic) with respect to LSB and HSB galaxies and correspondingly they have intermediate linewidth (twice the rotation rate).

\begin{figure}
\hspace{-1cm}\includegraphics[width=0.55 \textwidth]{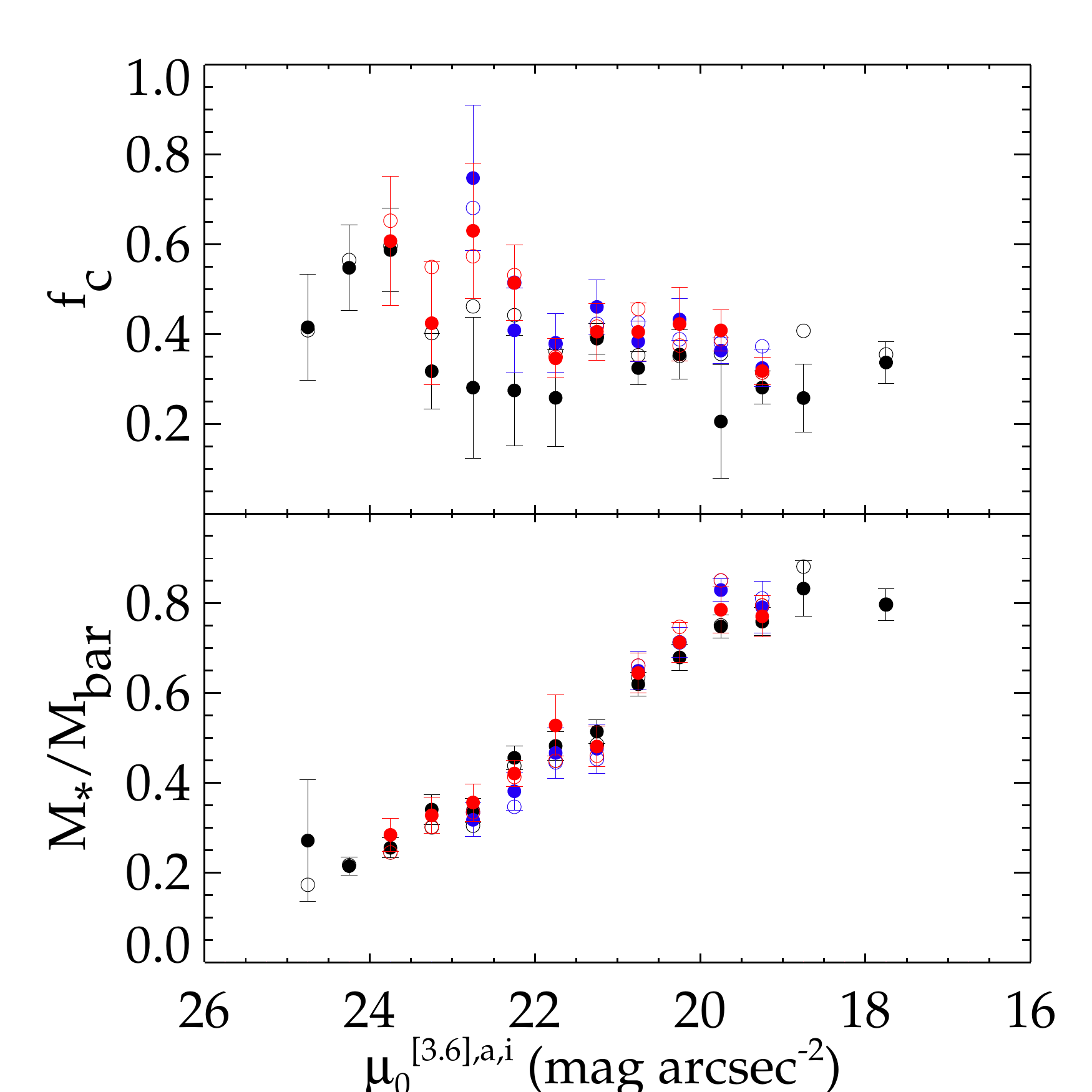}
\caption{Condensed baryonic fraction (top) and stellar to baryonic mass ratio (bottom) as a function of disc central surface brightness. Filled circles stand for mean values while open circles are median values. Bars stand for the standard error of the means. Black, red and blue colours stand for the entire sample (largest range of $\mu_0$), for the sample amputated of extreme inclination value galaxies (intermediate range of $\mu_0$) and in addition of very late or early type galaxies (smallest range of $\mu_0$) respectively. Mean and median are marked only if at least 5 galaxies contribute to the value computation.}
\label{fig:fraction}
\end{figure}

To go further into the baryonic Tully-Fisher relation, we follow \citet{2014AJ....147..134Z}: we consider a simple galaxy model where the mass of a halo of characteristic velocity, $v$, is $v^3$ \citep[$M~\propto~v^2~r$ and $r~\propto~v$, ][]{2010gfe..book.....M}. Then, if the baryonic Tully-Fisher relation holds, a straight line with a slope of 3 should fit the data. In hydrodynamical simulations, a slope of 3.4 seems in good agreement with simulated data \citep{2001ApJ...555..240B}. \citet{2014AJ....147..134Z} found a slope of 3.5$\pm$0.2 for observational S$^4$G data. The goal of this work is not to derive the best baryonic Tully-Fisher relation for the dataset used in this paper \citep[we refer the reader to][for that]{2014AJ....147..134Z} but to compare ISB position on the relation with respect to that of LSBs and HSBs. 

Accordingly, a linear relation of slope 3.4 and zero point adjusted with typical values of the Milky Way (linewidth or twice the rotation rate of $440\,\kms$ and baryonic mass of 8.3 10$^{10}$ M$_\odot$) is superimposed onto the data split into HSBs, LSBs and ISBs. The relation (solid line) is as expected in perfect agreement with HSBs on Figure \ref{fig:relation}. Varying the zero point while keeping the slope fixed only increases the scatter. 

On the contrary, the scatter is decreased for LSBs (dashed line) and ISBs (dotted line) when modifying the zero point (reduced for ISBs and increased for LSBs) while keeping the slope unchanged. Still, the scatter is quite large in both cases and certainly larger than that obtained for HSB galaxies, though this is to be expected since LSBs and ISBs have larger fractions of gas which do not correlate as nicely with linewidths as stellar masses do. This implies a posteriori that the Milky Way is a typical HSB while LSBs and ISBs with the same rotation rate as the Milky Way would have different baryonic masses.

\citet{2014AJ....147..134Z} noted that the baryonic content considered (${\rm M}_{\rm bar}$, the sum of the stellar and gas masses as defined by equation \ref{eq:mbaryonic}) accounts only for the condensed fraction of baryons onto the central region of spiral galaxies (both extremely cold and hot material and far from the centre baryons are not considered). Since the baryonic Tully-Fisher relation holds for typical spiral disc galaxies like the Milky Way with an estimated total mass of 1.2$\times10^{12}$ M$_\odot$, \citet{2014AJ....147..134Z} found this fraction to be a constant 40\% ($\frac{8.3 \times 10^{10}}{1.2 \times 10^{12}} \times \frac{1}{0.1649}$) assuming a universal baryon to total matter ratio of 0.1649 \citep[][]{2013ApJS..208...19H}. 

Assuming the same universal baryon to total matter ratio, it is possible to derive the fraction of condensed baryons as a function of disc central surface brightness. For typical spiral disc galaxies, we expect to find the 40\% condensed baryonic fraction. While keeping the slope fixed, we adjust the zero point of the baryonic Tully-Fisher relation to minimize the scatter for each galaxy subsample per bin size. The top panel of Figure \ref{fig:fraction} shows such condensed baryon fractions. The black colour represents the entire galaxy sample. A clear break in the 40\% expected fraction is visible at the gap value (22~mag~arcsec$^{-2}$). The 40\% fraction holds mostly for typical HSB galaxies. As for LSB galaxies, since they are mostly irregular galaxies, no specific baryonic fraction is expected. Excluding galaxies of extremely low or high inclinations (red colour) results in a higher confidence level in both linewidths and disc central surface brightnesses but does not alleviate the observation. Removing both very late and early types (keeping just the types between S0 and Sdm) tends to smooth the trend. Nevertheless the gap value corresponds to a change of regime of the value in the (restricted) samples. 

Next, the stellar to baryonic mass ratio is plotted as a function of disc central surface brightness in the bottom panel of Figure \ref{fig:fraction}. Again filled black dots stand for mean stellar to baryonic mass ratios given a disc central surface brightness (a bin size of 0.5~mag~arcsec$^{-2}$) while open black dots are medians of stellar to baryonic mass ratios. The trend is clear: the mean stellar to baryonic mass ratio increases from LSBs to HSBs through ISBs. There seems to be a break in the linear relation between stellar to baryonic mass ratio and $\mu_0$ around the gap value ($\sim$ 21-22 mag arcsec $^{-2}$). Keeping only galaxies with intermediate inclination values (red colour) does not change the observed trend. Discarding in addition galaxies of type earlier than S0 and later than Sdm (blue colour) leaves the observation unchanged. This observation is in agreement with the fact that galaxies in the gap are mostly a mix of exponential discs  with and without bars in the same proportion. Still individually, the left panel of Figure \ref{fig:relation} shows that ISB galaxies tend to have similar stellar and gas masses while LSBs have more gas and HSBs have higher stellar masses. 

\citet{2013MNRAS.433..751S} studied the impact of having a close neighbor ($d<80\;kpc$, $v\pm \,200\, \kms$) on $\mu_0$ as mergers / interactions could move galaxies into the gap separating the HSB and LSB galaxies \citep{1997ApJ...484..145T}. They found only a few galaxies with close neighbors with a quite flat $\mu_0$ distribution. With no particular trend revealed when looking at galaxies with close neighbors to explain the existence of ISBs with clear intermediate characteristics, we turn toward the global environment through the cosmic web in the next section.

\section{Determining the Environment with the Cosmic Web}

\begin{figure}
\includegraphics[width=0.5 \textwidth]{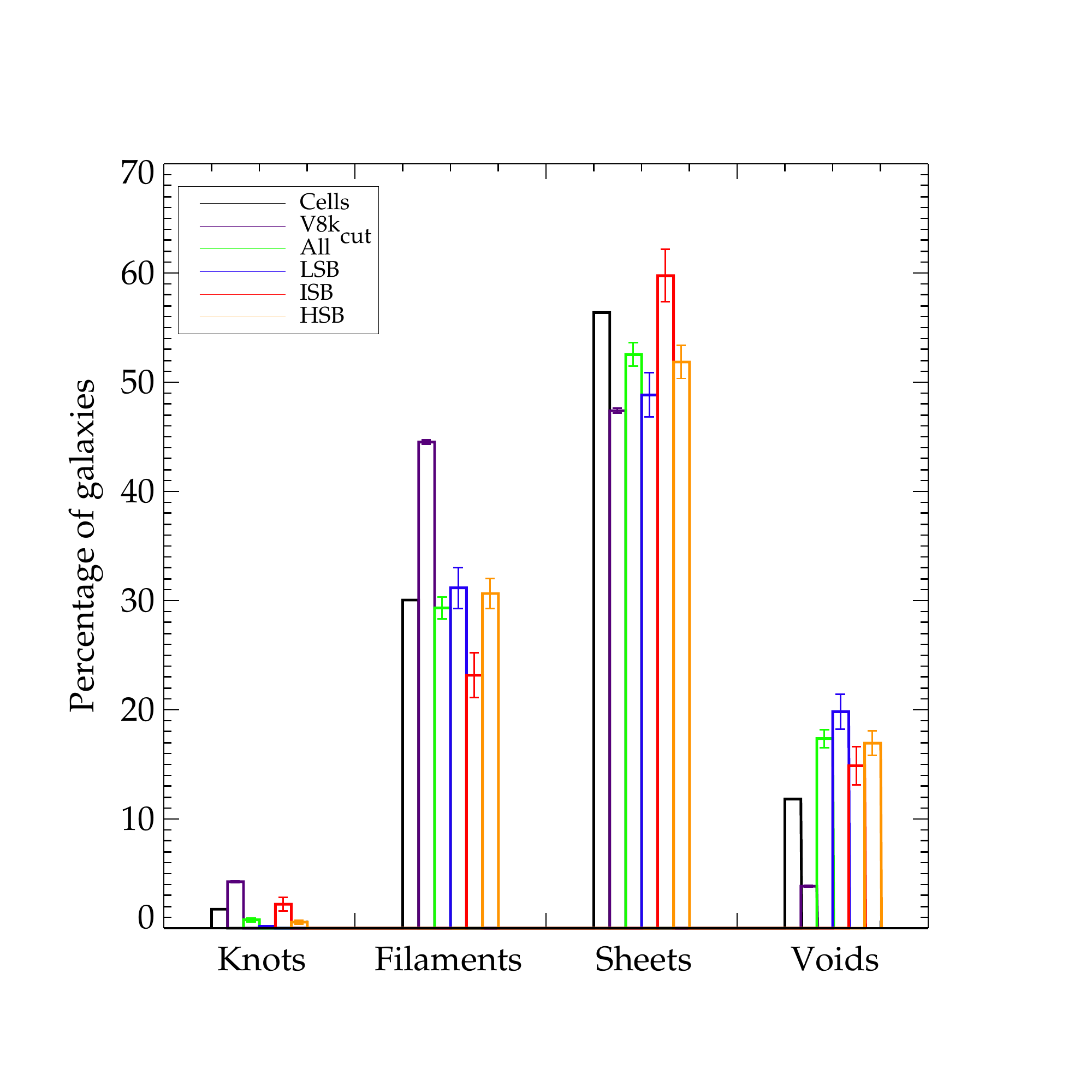}\\

\vspace{-2.5cm}
\includegraphics[width=0.5 \textwidth]{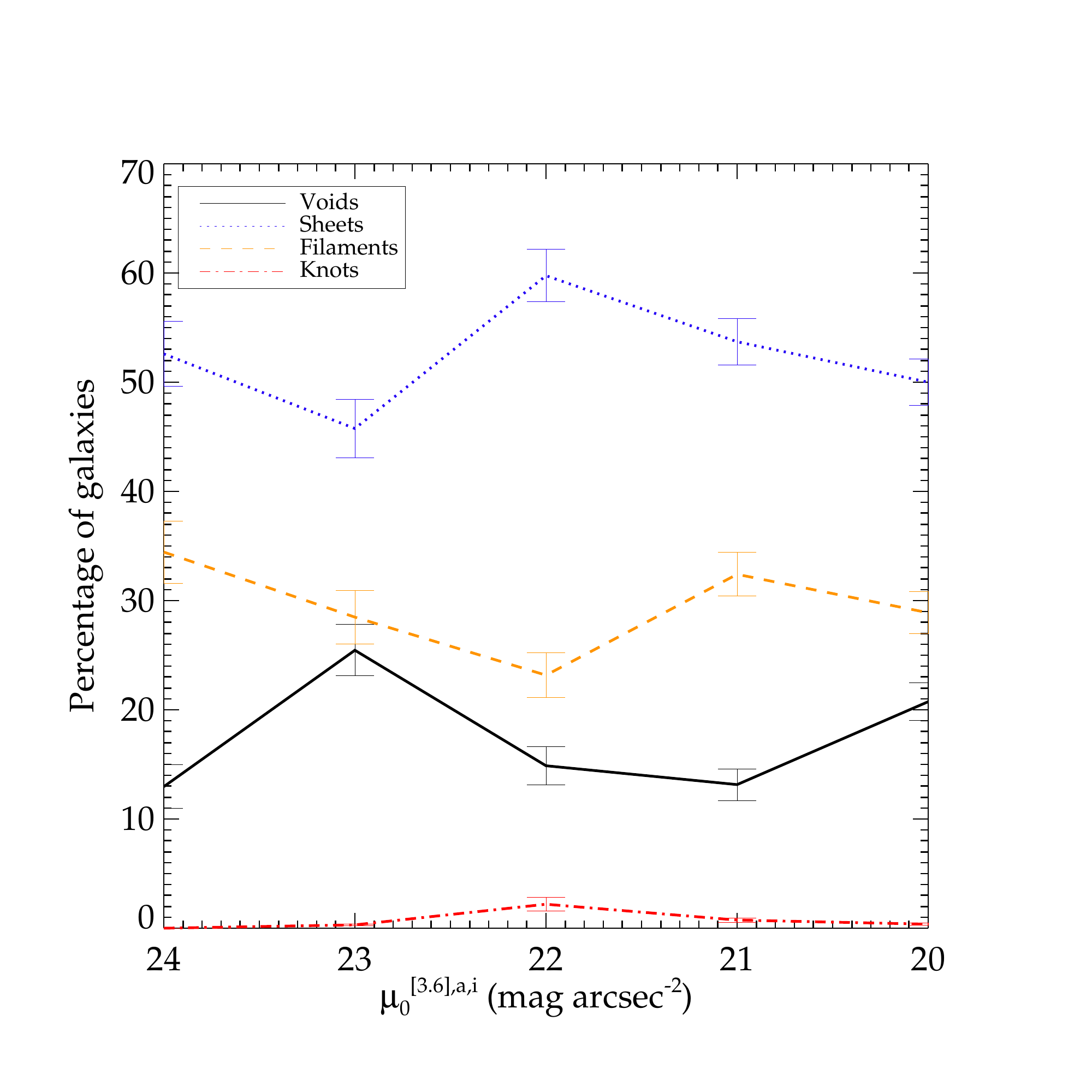}\\
\vspace{-1cm}

\caption{Top: Percentage of galaxies in knots, filaments, sheets and voids according to their category (LSBs, ISBs and HSBs and all), assuming a zero threshold. The distribution of cells of 1 \hMpc\ or repartition of the cosmic web into knots, filaments, sheets and voids is also given within 30 \hMpc\ (black line or first bar). Environments are given for galaxies from the V8k redshift catalogue cut at 30 \hMpc\ (violet line or second bar), for the selected S$^4$G sample (green line or third bar) removing both highly inclined and close to face-on galaxies. The same is done for the LSBs (blue line or fourth bar), for the ISBs (red line or fifth bar) and for the HSBs (orange line or sixth bar). Bottom: Percentage of galaxies in knots (red dot-dashed line), filaments (orange dashed line), sheets (blue dotted line) and voids (solid black line) as a function of their disc central surface brightness assuming a null threshold. ISBs have a higher probability to be in sheets than their counterparts.}
\label{fig:cosweb}
\end{figure}

In this section, we use the cosmic web to determine if ISBs exhibit a preference for their environment. To derive the cosmic web a number of methods have been developed over the years \citep[e.g.][among others]{2008MNRAS.383.1655S}. Here, we choose to characterize the cosmic web according to the Hessian formalism applied, not to the gravitational potential \citep[the result is the tidal tensor,][]{2007MNRAS.381...41H},  but to the velocity field \citep[this gives the shear tensor,][]{2012MNRAS.425.2049H,2012MNRAS.421L.137L}. Detailed equations are given in \citet{2014MNRAS.441.1974L} and we only remind here the adopted definition of the shear tensor $\Sigma$:
\begin{equation}
\Sigma_{ij}=-\frac{1}{2H(z)}(\frac{\partial v_i}{\partial r_j}+\frac{\partial v_j}{\partial r_i})
\end{equation}
where i,j are x,y or z and H(z) is the Hubble Constant as a function of the redshift, $v$ is the velocity field and $r$ the location vector.
Evaluation of eigenvalues of the velocity shear tensor permits the classification of volumes of the Universe with such a velocity field as either knots, filaments, sheets or voids. It is thus straightforward to determine the cosmic web environment of any given galaxy. The sample extends to $\sim$20 \hMpc, thus constrained simulations of the Local Universe \citep{2010arXiv1005.2687G} obtained with cosmicflows-1 \citep[catalog of radial peculiar velocity extending to 30 \hMpc, ][]{2008ApJ...676..184T} as described in \citet{2014MNRAS.437.3586S} are large enough to compute the necessary cosmic webs. The Large Scale Structure (LSS) of these simulations resemble the Local Universe within 30~\hMpc\ with an accuracy of a few megaparsecs (typically 2-3 \hMpc). We apply a cloud-in-cell to these different simulations to obtain a cell size of 1 \hMpc\ and we smooth at 1.6 \hMpc. The cosmic web is computed for the ten constrained simulations of the Local Universe obtained with cosmicflows-1 \citep[available from][]{2014MNRAS.437.3586S}. It is then possible to study the environments of the ISB galaxies in contrast to their counterparts LSB and HSB. Environmental types are determined by studying eigenvalues and their positions with respect to a threshold value. With a null threshold and the definition used in \citet{2012MNRAS.425.2049H} for the velocity tensor, three eigenvalues smaller than the zero threshold (i.e. three negative eigenvalues) correspond to a void while three positive values correspond to a knot. Two negative and one positive values constitute a sheet while the opposite stands for a filament. The nature of environments to which LSBs, HSBs and ISBs belong is then identified in the ten cosmic webs. Results are stable in the ten realizations and we do not see any reason to give preference to one of the simulations, consequently we average the results over the ten simulations for histograms and curves and take median values when considering galaxies individually. 

The distributions of galaxies for each environment, in knots, filaments, sheets and voids, are given in the histogram at the top of Figure~\ref{fig:cosweb}, for the sample excluding only extreme inclination galaxies as well as for LSB, ISB and HSB galaxies separately. For such a discrete set of data, the best way to estimate the statistical sampling error or a confidence interval, is a technique known as the Binomial proportion. This assumes that the distribution of errors about a binomially-distributed observation (namely, the web classification of each galaxy) is a normal distribution. The Binomial proportion then returns the one-sigma confidence in the fraction of galaxies assigned a specific web type. Because of the residual cosmic variance, the 10 realizations (although constrained) act as a smoothing factor on the variations of percentage. The mean one-sigma confidence intervals are divided by the square root of the number of realizations. These are shown as the error bars in Figure~\ref{fig:cosweb}. A similar process is conducted with galaxies from the V8k redshift catalogue cut at 30~\hMpc\ (available at EDD) and for the set of cells of 1 h$^{-1}$ Mpc constituting the simulation within 30~\hMpc. Histograms are also plotted in the same Figure and enable us to compare the results. Typically, there are expected to be approximately the same number of galaxies in filaments and sheets then fewer in knots and in voids \citep[e.g.][]{2015ApJ...799...45F}. Distributions shown on the Figure are in agreement with these trends. As for the repartition of cells, they depends more dramatically on the chosen threshold. We choose a zero threshold to distinguish between zones of collapse from zones of expansion, fractions obtained for the different environmental types are valid for this threshold. We will analyze later on the values of a threshold-independent parameter.

Clearly the selected S$^4$G sample has only a few galaxies in knots and from that perspective differs from the V8k catalogue which has 5 times more galaxies in knots. This is expected as galaxies selected for this paper are spiral late-type galaxies which do not reside preferentially in dense regions, in contrast to their counterparts \citep[early-type elliptical galaxies, e.g.][]{1980ApJ...236..351D,2004MNRAS.353..713K,2004MNRAS.348.1355B,2015MNRAS.446.1458M,2014MNRAS.445..988N}. They are also on average less numerous in filaments but more populous in voids, with respect to the V8k sample, which is also in agreement with expectations. The main trend is that ISBs have a higher probability to be in sheets than in filaments w.r.t. the whole sample (`All') whilst the opposite is true for the LSBs. The HSB galaxies have intermediate percentages, between those for ISBs and LSBs,  in both cases in agreement with average values obtained for the entire sample as expected as HSBs constitute the majority of the sample. This results in an excess of ISB galaxies at the level of 10 $\pm$ 5 \% and a deficit of LSB galaxies in sheets at approximately the same level while it is the opposite in filaments. The bottom of Figure \ref{fig:cosweb} confirms the cosmic classification trend as a function of the disc central surface brightness. Typically, using a null threshold, galaxies of intermediate disc central surface brightness (centered on 22 mag arcsec$^{-2}$) express a greater preference for sheets than their counterparts at higher or lower disc central surface brightness. Error bars have been derived as before.

As noted by \citet{1997ApJ...484..145T}, the source of the bimodality may be an instability when the density of dark matter and baryons are co-dominant in the centres of galaxies, i.e. that when the density of baryons is greater the galaxy has a tendency to become an HSB, and when the density is lower, it has a tendency to become an LSB. Co-dominance, and being an ISB, is an unstable intermediate state from which galaxies could evolve to either a HSB or a LSB. Analysis of the clustering of these galaxies suggests that it is the interaction of galaxies that trigger the transformation from the LSB to the HSB state \citep{1993AJ....106..530B,1994MNRAS.267..129M,2004A&A...422L...5R}, and consequently in the higher density environments the fraction of LSB should decline.

As can be seen in Figure~\ref{fig:cosweb}, our results only partially support such a view. In the voids, the LSB percentage is higher than the HSB one, and the ISB percentage is smallest, indeed suggesting LSB can persist in this low density environment. In the sheets, this is reversed, the HSB percentage is higher than the LSB one, and for ISBs it is highest, consistent with more interactions producing more ISBs (transition galaxies) and ultimately more HSBs. However, in the yet higher density environment of filaments, there are few ISBs and no shortage of LSBs, indeed their fraction in the filamentary environment is slightly higher than that of the HSBs. It appears that the sheet web-type is a `goldilocks' environment for the ISB galaxies, and clearly a more nuanced view is required.

\begin{figure}
\hspace{-0.8cm}\includegraphics[width=0.5 \textwidth]{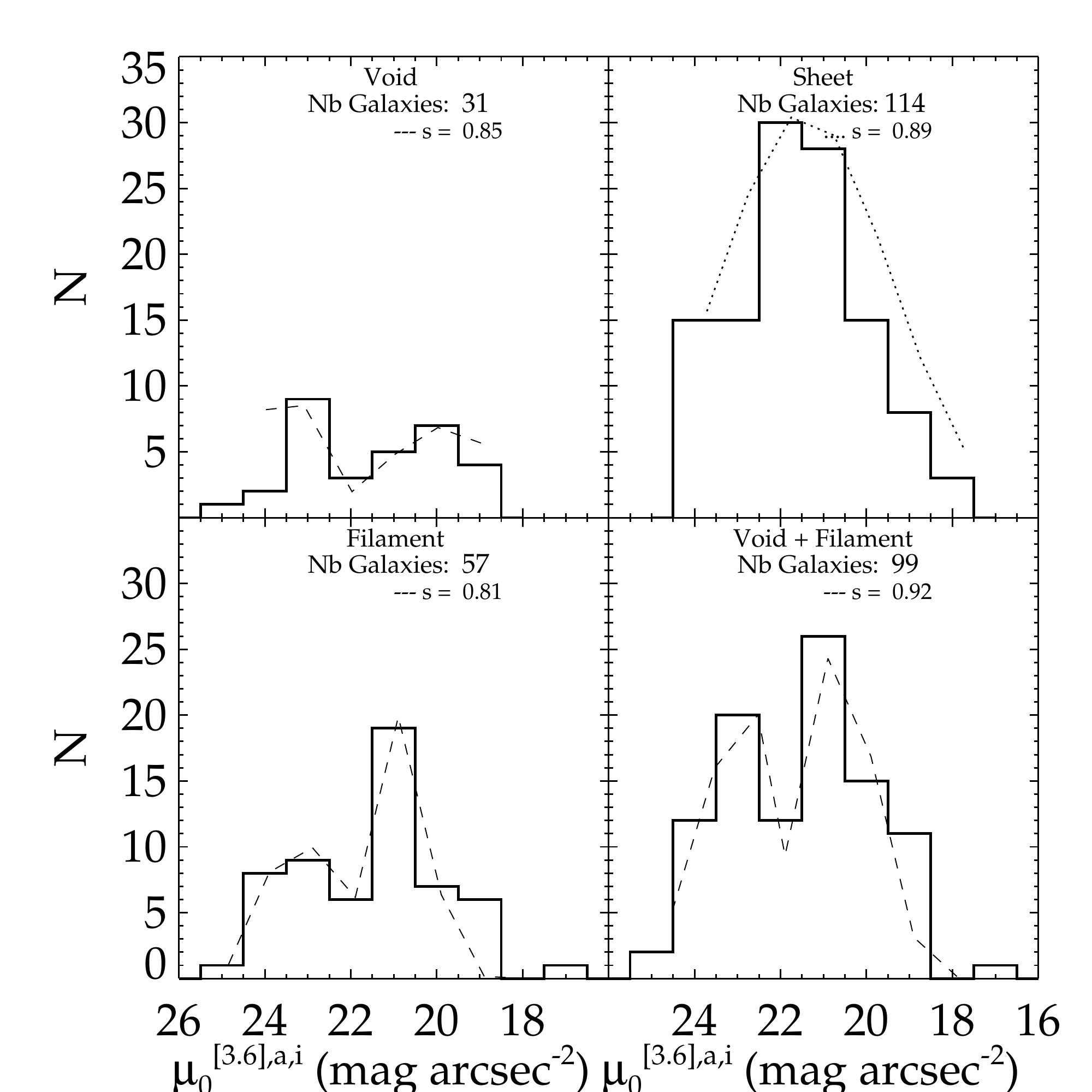}
\caption{Histograms of disc central surface brightness splitting galaxies according to their cosmic web environment. Significance levels (s) from F-test for a single or double Gaussian fitting are given. Clearly a double Gaussian fits well the distribution of disc central surface brightness in void, filament and void+filament while bimodality is absent in sheets (a double Gaussian cannot be fitted while a single Gaussian is in agreement with the distribution at the 89\% significance level).  }
\label{fig:cosweb1}
\end{figure}

To clarify this effect further we plot the distribution of central surface brightnesses in the different environments in Figure~\ref{fig:cosweb1}. In the sheet environment (top right panel) we can see that the bimodal distribution is completely absent, and an F-test gives a 89\% significance level between the variances of the data and of the single Gaussian least square fit, whilst the environments of voids, filaments and void+filament combination prefer the variance of a double-Gaussian at the 85, 81 and 92\% significance levels respectively. An additional sanity check reveals that for each double-Gaussian the residual between the two peaks is larger than twice their standard deviation. 

\begin{figure}
\hspace{-0.5cm}\includegraphics[width=0.55 \textwidth]{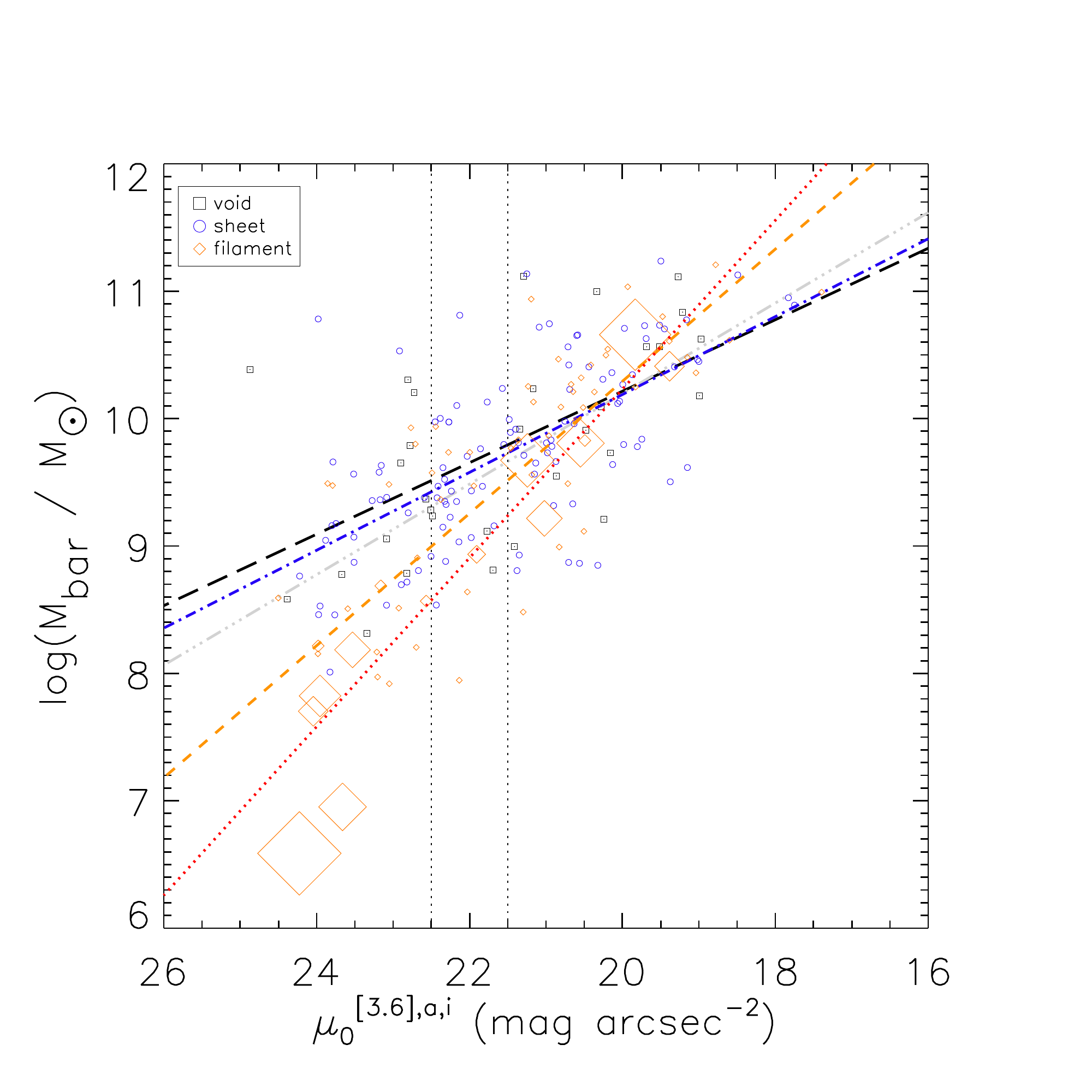}
\caption{Baryonic mass as a function of disc central surface brightness. Symbols represent galaxies, a black square for a galaxy in a void, a blue circle if it is in a sheet and an orange diamond when it is in a filament. The size of these latter is proportional to the number of close neighbor. The black dotted lines delimit the ISB zone. The coloured lines are linear regression fits selecting galaxies in voids (black dashed line), in sheets (blue dot-dashed line), in filaments (orange small dashed line) and in filaments with close neighbors (red dotted line). The grey triple dot-dashed line is a fit to all the points without distinction. Galaxies in filaments have the greatest baryonic mass range (larger slope). In sheets galaxies have more the same masses. LSBs are more massive in sheets than in filaments. LSBs with close neighbors strengthen this assertion. An opposite trend is observed for HSBs.}
\label{fig:cosweb2}
\end{figure}

In Figure~\ref{fig:cosweb2} we plot the baryonic mass vs. central surface brightness, additionally marking each with their median (over the different simulation, after checking that this plot is independent of the chosen realization) environment. When averaging other the different realization for each galaxy individually, the knot environment, already weakly represented, is mostly erased and thus we do not plot this environment type. Since galaxies in filaments may have nearby neighbors, we enlarge those symbols by the neighbor count, as found from the 2MASS catalogue \citep{2012ApJS..199...26H} taking a cut of 800~kpc and a velocity within $\pm 200 \, \rm km\,s^{-1}$.

Within Figure~\ref{fig:cosweb2} we can see the trend for higher central surface brightness galaxies to have higher baryonic mass, however perhaps more interesting is that for LSBs there is a split between high baryonic mass LSBs which are mostly in sheets and voids and low baryonic mass LSBs which are mostly in filaments. For HSBs there is little evidence for this, as baryonic masses in the different environments appear to overlap. In order to capture this trend we have additionally fit linear regressions between baryonic mass and disc central surface brightness for the environments separately and as a whole. 

These trends reveal that for the galaxies in filaments, and in particular for those with nearby neighbors, the baryonic mass is a much stronger function of central surface brightness than for the void and sheet environments.
In sheets, the mass increases weakly with the disc central surface brightness, while in filaments (with additionally close neighbors), the baryonic mass increases almost (more than) twice as fast. The residual between the slopes (voids and sheets versus filaments) is about 0.3 in absolute value and the uncertainty on the fits are about 0.1 each, the significance level of the difference in strength of the correlations is 2.5 (2) $\sigma$.
This suggests that we are seeing an effect of satellites, as the star-formation rates of satellites in groups are known to be affected (e.g. \citealp{2002MNRAS.334..673L} perhaps extending to loose groups, \citealp{2015ApJ...800L...4C}), and satellites are more common in the filamentary environment \citep[e.g.][]{2015ApJ...800..112G}, i.e. whilst the star formation in the filamentary environment is in general higher \citep[e.g.][]{2014ApJ...796...51D}, this does not happen indiscriminately, the central objects undergoing much more accretion and star-formation than the satellites, which could become the low baryonic mass LSBs.

This provides an additional context for the distributions seen in Figures~\ref{fig:cosweb}~\&~\ref{fig:cosweb1}. Whilst LSBs may be promoted to HSBs by the interactions in the higher density environments, there can be a competing effect that the nature of accretion is becoming less stochastic in those same environments. More specifically, the HSB galaxies are primarily accreting the baryonic material, whilst the LSBs are not, setting up the stronger trend of central surface brightness to baryonic mass. 
Since accretion of material into the central regions can perturb the baryonic to dark ratio and thus between the LSB, ISB and HSB states, an implication of this `ordered' accretion may be to reduce the transitions overall, allowing the ISB fraction to fall and the distribution to relax back to bimodality. The sheet environment may then provide a local maxima for the transitions and consequently the (relative) ISB percentage: in the void environment there are insufficient galaxy interactions, whilst in the filamentary environment the accretion modes for high and low central surface brightness galaxies are sufficiently distinct that they are less likely perturbed into the alternative regime.

\begin{figure}
\hspace{-0.5cm}\includegraphics[width=0.55 \textwidth]{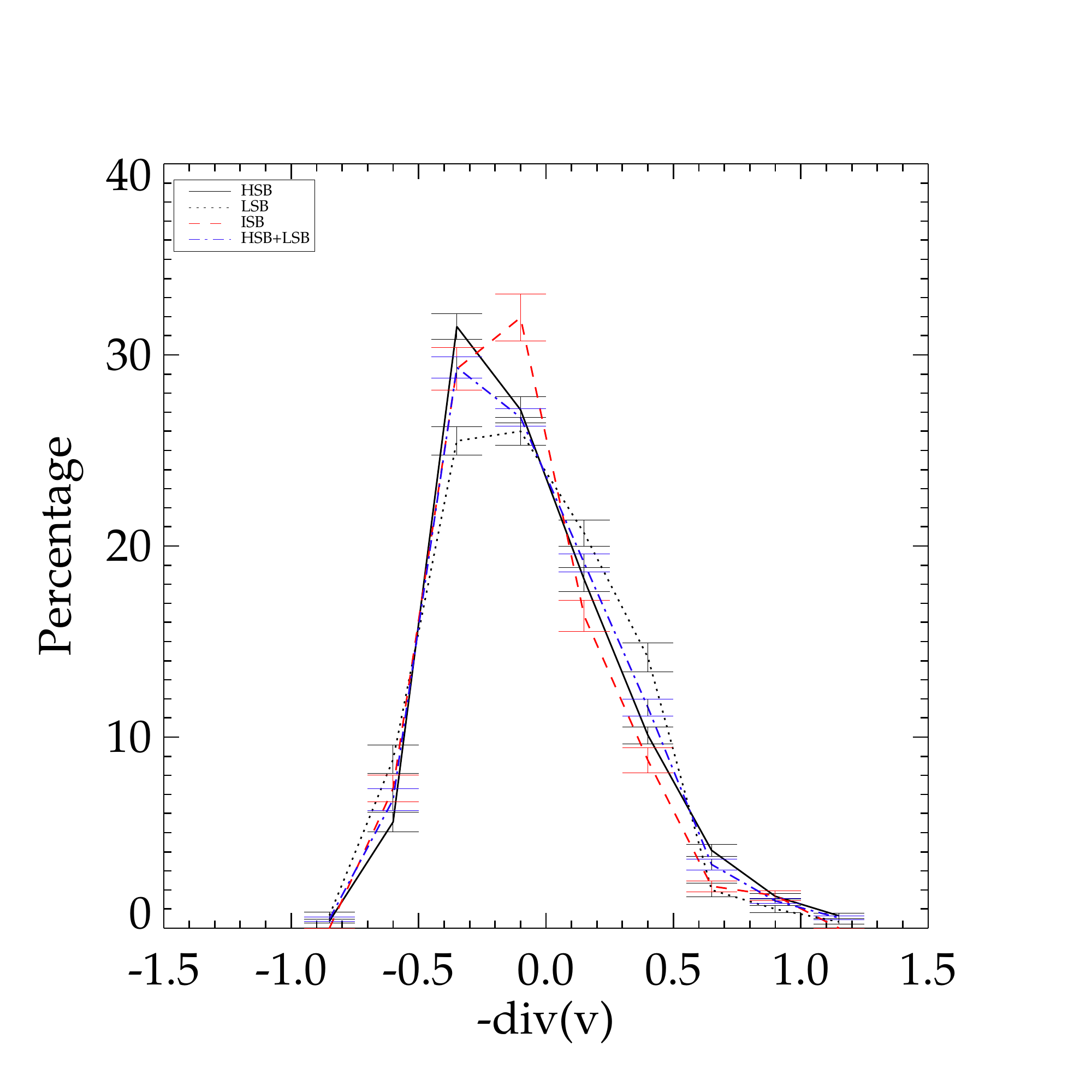}
\vspace{-1cm}
\caption{Values of the trace of the shear tensor at the positions of the LSBs, ISBs, HSBs and LSBs+HSBs. The trace of the shear tensor is the negative of the velocity divergence at the studied position. Their mean distribution for ISB galaxies peaks at $\approx$ -0.1 while those for LSB and HSB peak at $\approx$ -0.4, typically ISBs are in flows of lower expansion than their HSB and LSB counterparts.}
\label{fig:trace}
\end{figure}

As we mentioned however, the classification in different cosmic web types depend on the used threshold. We used the simple value of zero to distinguish  between regions of collapse and regions of expansion and doing so we find that ISBs are more prevalent in sheets than in filaments or even voids. We try to find a threshold-independent definition for the environmental preference of ISBs. In Figure \ref{fig:trace} we look at a threshold-independent parameter to confirm the particularity of ISBs: the trace of the shear tensor evaluated at the positions of the galaxies is the negative of the velocity divergence. The mean standard deviation for these values in the different simulations is about 0.2 thus a bin size of 0.25 is used to plot their mean distribution.
ÕKolmogorov- Smirnov test shows that distributions obtained for HSBs, LSBs and ISBs considering all the simulations together have a 75\% probability to differ. The distribution obtained for LSBs and that derived for ISBs are found to have the highest probability (99.3\%) to be different. Comparing together LSBs and HSBs versus ISBs, the probability that they differ is 95\%.
The distribution for ISBs is found to have a closer to zero mean (and peak) than that of the LSB and HSB galaxies ($\approx$ -0.1 against -0.4), indicating that on average ISBs reside in domains with less expansion (divergence) than the LSBs and HSBs. The distribution is also somewhat narrower for ISBs, so in addition at large scale negative divergences (approximately greater than 0) the wings of the LSBs and HSBs dominate, namely in domains of contraction the LSBs and HSBs are also dominant. In this way weak contraction (rather than strong expansion or contraction) is also a `goldilocksÕ regime for ISBs, which is in rough agreement with expectations that ISB galaxies tend to lie in less active environment (like sheets) rather than in filaments and voids.

\section{Conclusion}
While extreme galaxy parameters are always scrutinized to test theoretical models of galaxy formation and evolution, intermediate parameter values are often ignored, despite their important role in verifying our assumptions. This is the case of intermediate surface brightness galaxies, an under-represented population in the Universe. 
In this paper, intermediate surface brightness galaxies have been studied more closely. First the dichotomy in disc central surface brightness in a complete sample down to a magnitude limit of -16 in the B band of galaxies from the Spitzer Survey of Stellar Structures in Galaxies has been confirmed, using values obtained with a 2D decomposition, at the 95\% significance level (F-Test). Then stellar, gaseous and baryonic masses have been derived for ISBs as well as for their LSB and HSB counterparts. Intermediate surface brightness galaxies always reveal themselves to have intermediate values and characteristics. Looking at the stellar to baryonic mass ratio, they are again intermediate. This intermediate value could be due to the fact that they are a mix of exponential disc with and without bars, and bars trigger star formation in the central parts of galaxies. 
Still, when examined individually, most intermediate disc central surface brightness galaxies have as much gas as luminous material excluding the average as the sole cause of the observation. There is, however, a clear break in the $40\%$ condensed baryon fraction at their level in the distribution of mean condensed baryon fraction versus disc central surface brightness, although this break is alleviated by removing close to face-on galaxies (reducing error in linewidth and therefore position on the baryonic Tully-Fisher relation) and highly inclined galaxies (reducing the uncertainty on disc central surface brightness). That they have intermediate values indicate that they could be an unstable transitional stage between LSBs and HSBs. 

As for their preferred environment, taking a null threshold to discriminate between regions of collapse and regions of expansion, they are 5-$10\%$ more likely to be found in sheets and 5-$10\%$ less likely to be found in filaments when compared with LSBs and HSBs. Among all the parameters studied in this paper, two things in which they seem to be exceptional is their number and their repartition in the dynamical environment. They are under-populous in comparisons with LSBs and HSBs in general but more specifically in filaments, while they tend to gather in sheets. This leads to a bimodal distribution of disc central surface brightness in filaments, voids but not in sheets. In addition we find that the strength of the correlation between disc central surface brightness and baryonic mass depends on the environment. In sheets, the mass increases weakly with decreasing disc central surface brightness, while in filaments (with additionally close neighbors), the baryonic mass increases almost (more than) twice as fast at the 2.5 (2) $\sigma$ significance level. Because the classification in terms of cosmic web types depends on the threshold, we look also at the threshold-independent trace of the shear tensor (velocity divergence). Using Kolmogorov-Smirnov tests, we find that velocity divergence distributions differ at least at the 50\% significance level indicating that: ISB galaxies tend indeed to be in flows of lower expansion/collapse  while LSBs and HSBs are more present in flows of higher expansion/collapse at the 1$\sigma$ confidence level. 
 In the traditional picture ISBs are seen to be an unstable transition state between LSBs and HSBs, and less likely to form in the more isolated environments where the interactions are uncommon. We find, however, that in the highest density environments the LSBs are likely to be satellite galaxies, whose accretion is sufficiently constrained that it fails to promote them to HSBs, making the intermediate environment the `goldilocks' zone which maximally encourages the formation of ISBs.

After all these considerations, it seems obvious that ISB galaxies have intermediate characteristics and that their disc central surface brightness values is a crucial turnover in galaxy characteristics. Instability is the probable cause of their smaller statistical number thus they cannot survive long in environments with numerous interactions (large inflow activity) or they cannot form in environments without a sufficient amount of interactions (large outflow activity). There are still a number of open questions to explain the observed galaxy disc central surface brightness distribution. For sure ISBs are present in simulations where they are found to have as much dark as baryonic matter in their centre while HSBs have more baryonic matter in their centre and LSBs are dominated by dark matter at all radii \citep{2013MNRAS.434.3142A}. The number of such simulated galaxies is however too low to draw any statistical conclusion or any conclusion on whether their bimodal characteristic $\mu_0$ distribution is reproduced by simulations. Numerous questions that could perhaps be looked at more thoroughly with recent large hydrodynamical simulations such as Evolution and Assembly of GaLaxies and their Environments \citep[EAGLE, e.g.][]{2015MNRAS.446..521S} although the limitation in subgrid physics might not permit to retrieve results comparable with those of \citet{2013MNRAS.434.3142A}. 

\section*{Acknowledgements}
We thank the S$^4$G team for sharing their data with the community. We thank the anonymous referee whose useful comments helped improved the manuscript. JS acknowledges support from the Alexander von Humboldt Foundation. PC acknowledges support from the Leibniz Gemeinschaft through grant SAW-2012-AIP-5 129. NIL is supported by the Deutsche Forschungs Gemeinschaft.


\bibliographystyle{mnras}

\bibliography{biblicomplete}
\label{lastpage}
\end{document}